\documentclass[aps,pre,superscriptaddress]{revtex4}
\usepackage[usenames,dvipsnames]{color}

\usepackage{graphicx,color}
\usepackage{bm}
\usepackage{amsmath}
\usepackage{verbatim}
\usepackage{amssymb}
\usepackage{chemarr}

\newcommand{\bra}[1]{\langle{#1}|}
\newcommand{\ket}[1]{|{#1}\rangle{}}
\newcommand{\avg}[1]{\langle{#1}\rangle}

\newcommand{\beqn}{\begin{eqnarray}}
\newcommand{\eeqn}{\end{eqnarray}}
\newcommand{\beq}{\begin{equation}}
\newcommand{\eeq}{\end{equation}}
\newcommand{\abs}[1]{|#1|}

\newcommand{\mk}[2]{\newcommand{#1}{#2}}
\newcommand{\rmk}[2]{\renewcommand{#1}{#2}}
\newcommand{\ip}[2]{\langle#1|#2\rangle}

\renewcommand{\>}{\rangle}

\mk{\bbn}{b^+_nb^-_n}
\mk{\bbm}{b^+_mb^-_m}
\mk{\bbnn}{b^+_nb^-_n(n)}
\mk{\bbmn}{b_m^+b_m^-(n)}
\mk{\bb}{\bp_n\bl_n}
\mk{\bbbn}{\bp_n{\bar{b}^-_n}}
\mk{\bbbm}{\bp_m{\bar{b}^-_m}}
\mk{\blb}{{\bar{\bl_m}}}
\mk{\bG}{{\bf{{G}}}}
\mk{\bln}{\hat{b}_m^-(n)}
\mk{\bnn}{\hat{b}_n^-(n)}
\mk{\bmn}{\hat{b}_m^-(n)}
\mk{\bbarn}{\hat{\bar{b}}_n^-}
\mk{\bbarm}{\hat{\bar{b}}_m^-}
\mk{\betan}{\hat{\beta}^-_n}
\mk{\betam}{\hat{\beta}^-_m}
\mk{\bbar}{\bar{b}^-}
\mk{\bp}{\hat{b}^+}
\mk{\bd}{\hat{b}^+}
\mk{\bpk}{\hat{b}^{+k}}
\mk{\ad}{\hat{a}^+}
\mk{\adN}{\left(\hat{a}^{+}\right)^N}
\mk{\bl}{\hat{b}^-}
\mk{\al}{\hat{a}^-}
\mk{\qn}{q(n)}
\mk{\gn}{g(n)}
\mk{\qh}{\hat{q}(n)}
\mk{\gh}{\hat{g}(n)}
\mk{\qb}{\bar{q}}
\mk{\gb}{\bar{g}}
\mk{\Dn}{\hat{\Delta}_n}
\mk{\Gah}{\hat{\Gamma}(n)}
\mk{\Lah}{\hat{\Lambda}(n)}
\mk{\Deh}{\hat{\Delta}(n)}
\mk{\Gjk}{G^{jk}}
\mk{\bj}{\<j|}
\mk{\bk}{\<k|}
\mk{\Gl}{{G_\ell}}
\mk{\Del}{\Delta}

\mk{\ehat}{\bf{\hat{ e}}}
\rmk{\L}{{\cal L}}
\rmk{\H}{\hat{H}}
\mk{\Om}{\hat{\Omega}}
\mk{\n}{\langle n\rangle}
\mk{\Lop}{\hat{\cal L}}
\rmk{\l}{\ell}
\newcommand{\choo}[2]{\begin{pmatrix} #1 \\ #2 \end{pmatrix}}

\begin{document}

\title{Analytic methods for modeling stochastic regulatory networks}
\author{Aleksandra M. Walczak}
 \email{awalczak@princeton.edu}
\affiliation{Princeton Center for Theoretical Science, Princeton University, Princeton, NJ 08544}
\author{Andrew Mugler}
 \email{ajm2121@columbia.edu}
 \affiliation{Department of Physics, Columbia University, New York, NY 10027}
\author{Chris H. Wiggins}
 \affiliation{Department of Applied Physics and Applied Mathematics, Columbia University, New York, NY 10027}

\date{\today}
\begin{abstract}
The past decade has seen a revived interest in the
unavoidable or {\it intrinsic} noise in biochemical and genetic
networks arising from the finite copy number of the participating
species. That is, rather than modeling regulatory networks in
terms of the deterministic dynamics of concentrations, we model
the dynamics of the probability of a given copy number of the
reactants in single cells. Most of the modeling activity of
the last decade has centered on stochastic simulation of
individual realizations, i.e., Monte-Carlo methods for generating
stochastic time series. Here we review the mathematical
description in terms of probability distributions, introducing the
relevant derivations and illustrating several cases for which
analytic progress can be made either instead of or before turning
to numerical computation.
\end{abstract}
\maketitle

\tableofcontents

\section{Introduction}
Cellular
processes rely on biochemical signaling,
i.e.\
chemical interactions
among individual
molecules.
Theoretical descriptions of biochemical
processes rely on considering the concentration changes of the molecules involved in the reactions, and specifying the types of regulatory interactions between them. The theoretical tools used to describe many types of biochemical reactions share many similarities. In this chapter we give an overview of the analytical approaches to study biochemical kinetics using the example of small gene regulatory networks. 

The regulation of genes by transcription factor proteins is an intrinsically stochastic process, owing to the small numbers of copies of molecules involved. 
With
the development of imaging techniques in molecular biology, we are able to observe directly the fluctuations in the concentrations of proteins and mRNAs and, by measuring the intensity profiles of fluorescence markers, measure full probability distributions \cite{Elowitz, Ozbudak, Pedraza, Raj}. Experiments over the last decade have shown that in fact gene regulation is a noisy process, and noise can propagate in gene networks. 

Many methods for solving the resulting
stochastic
equations rely on computer simulations. The efficiency of these methods has been greatly advanced in the last several years \cite{zontenwolde05_a, zontenwolde05, allentenwolde06, valerianiallentenwolde07, morelliallentenwolde08, munskykhammash06, munskykhammash07,samadpetzgold05, lampoudigillespie09,petzgoldgillespie09, chevaliersamad09, verena09}
. However numerical simulations are naturally limited to a specific choice of parameters, and changing the parameters requires a completely new calculation. Furthermore they suffer from the curse of dimensionality:
the computational runtime grows prohibitively as the number of species increases
These problems can be bypassed by developing analytical approaches, which often require certain approximations.

This chapter is intended as a tutorial on theoretical descriptions of biochemical kinetics. For clarity of exposition, we start by introducing the simplified kinetic description of the production of proteins. At the level of molecular kinetics, this is the simplest type of process that can occur. We first consider the deterministic description of the system and then introduce noise. After familiarizing the reader with different levels of description, we discuss models of regulation. We also present a wide spectrum of analytic tools used to find the steady state probability distributions of protein copy numbers in the cell. We point out that
while
for concreteness we focus on the case of gene regulation,
the methods presented in this chapter are very general. 

The results presented in this review are not new; some can be found in textbooks \cite{vanKampen, Gardiner, Zwanzig}, while others have been derived more recently in the context of gene regulation \cite{KeplerElston, Swain,hasty2,  hasty3, bialekadv01, mehta08, Thattai2, Hornos, mtv, cspan, Mugler, bialek2005physical, tkacikwalczakbialek09, tkacikwalczakbialek10, IyerBiswas, Raj, warrensorintenwolde06, PhysRevLett.97.068102,Paulsson, elfehrenberg}.
The goal of this review is to give the reader who is not familiar with analytical methods for modeling stochastic gene regulation an overview of the mathematical tools used in the field. Naturally, we are not able to cover all the developments in the field, but we hope to give the reader a useful starting point. For concreteness we also limit our discussion to the case of small gene networks and do not discuss approximations used to describe larger networks, which is currently an active area of research in many communities \cite{mtv, cspan, verena09,chevaliersamad09, munskykhammash06, munskykhammash07}.

\section{Simple birth--death process}\label{BD}
In this section we describe a simple birth death process for one species, absent of regulation. We first remind the reader of the deterministic description given by chemical kinetics. Next we introduce the full probabilistic description (given by the master equation), and, restricting our discussion to the simplest case, show how the deterministic equations arise as the dynamics of the mean. After calculating the variance and, we introduce the generating function formalism, useful in solving the master equation. This formalism may also be described using raising and lowering operators familiar to physicists from quantum mechanics. Finally we relate this description to the Fokker-Plank equation, the analogous master equation for continuous state variables (e.g., real-valued coordinates rather than the integer-valued copy number). These results will be used in later sections when we introduce autoregulation and regulation among different species. 

\subsection{Deterministic description: the kinetic rate equation}
In the simplest case,
the number of copies $n$ 
of a protein species $X$
can change either due to the production of a protein, which occurs at a constant rate $\tilde{g}$, or due to degradation of a protein, which occurs at a constant rate $r$:
\begin{equation}
\emptyset \xrightleftharpoons[r]{\tilde{g}} X.
\end{equation}
Here we condense the complicated molecular machinery of transcription, translation, and protein modification
into a single constant production rate.
Similarly we do not specify the molecular processes which have led to degradation; for simplicity we imagine either dilution due to cell division or active degradation with a constant rate.

When $n$ is large, the dynamics for the mean $\<n\>$ are well approximated by the continuous, deterministic description provided by the kinetic rate equation for the concentration $c\equiv\<n\>/V$ in a volume $V$:
\beq
\frac{d c}{d t}=\frac{\tilde{g}}{V}- rc.
\label{kinone}
\eeq
The solution of Eqn.\  \ref{kinone} is
\beq
\label{nt}
c(t)=\frac{\tilde{g}}{Vr}+e^{-rt} \left[c(0)-\frac{\tilde{g}}{Vr}\right],
\eeq
where $c(0)$ is the concentration of proteins at initial time $t=0$. In steady state
the mean number of proteins  is simply the ratio of the production and degradation rates, $\avg{n}=cV=\tilde{g}/r$, which is easily seen by taking either $d c/dt =0$ in Eqn.\ \ref{kinone} or $t \rightarrow \infty$ in Eqn.\ \ref{nt}.

\subsection{Introducing noise: the master equation}
The general probabilistic description of chemical reactions, respecting the finite copy number of the reactants, is the master equation, which specifies the rate of change of $p_{\bf n}$, the probability that there are ${\bf n}=(n_1,n_2,\ldots,n_L)$ copies of the $L$ reactants. The macroscopic, deterministic description in terms of chemical kinetics is recovered by considering the dynamics for the mean concentration of these reactants, i.e.~the vector $\avg{{\bf n}}/V=(\avg{n_1}/V, \avg{n_2}/V, \ldots \avg{n_L}/V)=(c_1,c_2,\ldots c_L)$. Such a description in terms of a summary statistic necessarily ignores a great amount of information, including all information about fluctuations about these mean values.

The general form of the master equation (c.f.~Appendix \ \ref{app:master}), set by conservation of total probability, is
\begin{equation}
\label{eq:mastergeneral}
\dot{p}_{\bf n}= \sum_{{\bf n'}} \left[ w_{{\bf n}{\bf n'}}(t) p_{\bf n'} - w_{{\bf n}{\bf n'}}(t) p_{\bf n} \right],
\end{equation}
For the case of only one species, we need specify only the dynamics of ${\bf n}=n$. Restricting
further to the case of a simple birth-death process, there are only two nonzero contributions from the transition matrices $w_{nn'}$,
given by
\beqn
w_{n+1,n}(t) &=& \tilde{g}, \\
w_{n-1,n}(t) &=& rn,
\eeqn
all other values of $w_{nn'}$ being 0. Under this restriction Eqn. \ \ref{eq:mastergeneral} reduces to the familiar
\begin{equation}
\label{eq:master0}
\dot{p}_n = -\tilde{g} p_n - rnp_n + \tilde{g} p_{n-1} + r(n+1)p_{n+1}.
\end{equation}
Qualitatively, the four terms on the right hand side represent
how the probability of having $n$ proteins can either (i) decrease in time, if there are $n$ initially and one is either produced (first term) or degraded (second term), or (ii) increase in time, if there are either $n-1$ initially and one is produced (third term) or $n+1$ initially and one is degraded (fourth term).

The dynamics of the mean number of proteins  $\avg{n} = \sum_{n=0}^\infty n p_n$ are readily obtained (see Appendix \ref{app:summing}), as
\beq
\label{avgn}
\frac{d \avg{n}}{dt} = \tilde{g} - r\avg{n}.
\eeq
Eqn.\ \ref{avgn} shows that the dynamics of the mean of the protein distribution reproduces the kinetic rate equation, Eqn.\ \ref{kinone}.

From here on, we will re-scale time $t$ by the degradation rate $r$, such that $rt \rightarrow t$,
and define $g=\tilde{g}/r$, making the master equation
\begin{equation}
\label{eq:master}
\dot{p}_n = -g p_n - np_n +g p_{n-1} + (n+1)p_{n+1}.
\end{equation}

\subsubsection{
Steady state solution}\label{steadystateBD}
The master equation can be rewritten in terms of shift operators $E^+$ and $E^-$ which increase and decrease the number of proteins
by one, respectively
\cite{vanKampen}, i.e.\
\beqn
E^+f_n &=& f_{n+1}, \\
E^-f_n &=& f_{n-1},
\eeqn
for any function $f_n$.
We begin by writing Eqn.\ \ref{eq:master} in terms of only $E^+$ to make clear that, as is the case for any one-dimensional master equation, its steady state can be found iteratively.  In terms of $E^+$,
\beq
\label{eq:master_op}
\dot{p}_n = (E^{+}-1)(n p_{n}-g p_{n-1} ).
\end{equation}
Setting $\dot{p}_n=0$ for steady state requires that the second term in parentheses vanish, giving for $n>0$ the recursive relation
\beq
\label{recurse}
p_{n}=p_{n-1} \frac{g}{n}.
\eeq
Starting with $n=1$ and computing the first few terms reveals the pattern
\beq
\label{recurse2}
p_{n}= \frac{g^n}{n!} p_0
\eeq
where $p_0 = e^{-g}$ is set by normalization:
\begin{equation}
1 = \sum_{n=0}^\infty p_n = p_0 \sum_{n=0}^\infty \frac{g^n}{n!} = p_0 e^g.
\end{equation}
Thus the steady-state probability of having $n$ proteins is the Poisson distribution,
\begin{equation}
\label{eq:pstar1}
p_n = \frac{g^n}{n!}e^{-g},
\end{equation}
with parameter $g$, the ratio of production to degradation rates.

We remind the reader that the variance $\sigma^2 = \avg{n^2}-\avg{n}^2$ of the Poisson distribution is equal to its mean $\avg{n}$ (Appendix \ref{Appendix_meansofBD}).  Therefore the standard deviation $\sigma$ over the mean falls off like
\beq
\frac{\sigma}{\avg{n}} = \frac{\sqrt{\avg{n}}}{\avg{n}} = \frac{1}{\sqrt{\avg{n}}},
\eeq
demonstrating that the relative effect of fluctuations diminishes for large protein number.
In Appendix \ref{app:gauss} we show that in the limit of large protein number the steady state asymptotes to a Gaussian distribution with mean and variance equal to $g$.

For comparison with the other representations of the master equation described below, we now write Eqn.\ \ref{eq:master} in terms of both shift operators $E^+$ and $E^-$:
\beq
\label{eq:master_op2}
\dot{p}_n = (E^+-1)(n-gE^-)p_n.
\eeq
Eqn.\ \ref{eq:master_op} can be rewritten slightly by inserting between the parenthetic terms the unit operator $1 = E^-E^+$ and distributing the $E^-$ to the left and the $E^+$ to the right, giving
\beq
\label{eq:master_op3}
\dot{p}_n = -(E^--1)[(n+1)E^+-g]p_n.
\eeq
where the negative sign has been factored out of the first parenthetic term.

\subsubsection{
Generating function representation}
\label{sec:Gx}

Not all master equations are solvable by straightforward iteration.  A more generalizable way to solve
a master equation is by the introduction of a generating function \cite{vanKampen}.
Here we demonstrate the generating function approach on the birth--death master equation, Eqn.\ \ref{eq:master}.

The generating function $G$ is defined as
\begin{equation}
\label{eq:Gxdef}
G(x,t) \equiv \sum_{n=0}^\infty p_n(t) x^n,
\end{equation}
a power-series expansion in a continuous variable $x$ whose coefficients are the probabilities $p_n$.  Since $x$ is a variable we introduce, we note that defining $x = e^{i\kappa}$ makes clear that the generating function is equivalent to the Fourier transform of $p_n$ in protein number (this point is further developed in Appendix \ref{app:nbasis}).
The probability distribution may be recovered via the inverse transform
\beq
\label{invtrans}
p_n(t)=\frac{1}{n!} \partial_x^n G(x,t)|_{x=0}.
\eeq
Additionally we note that the $\ell$th moment may be generated by
\beq
\avg{n^\ell}=\partial_x^nG|_{x=1},
\eeq
(which is the reason for the generating function's name).

The utility of the generating function for solving a master equation is that it turns an infinite set of ordinary differential equations (e.g.\ Eqn.\ \ref{eq:master}) into a single partial differential equation.  
To see this  here we multiply Eqn.\ \ref{eq:master} by $x^n$ and sum over $n$ to obtain (see Appendix \ref{app:summing} for the detailed derivation):
\beq
\label{eq:Gss}
\dot{G}  = -(x-1)(\partial_x-g) G.
\eeq
We see immediately from comparison of Eqns.\ \ref{eq:master_op3} and \ref{eq:Gss} that the representations of operators $E^-$ and $(n+1)E^+$ in $x$ space are $x$ and $\partial_x$, respectively. 

In steady state ($\dot{G} = 0$) Eqn.\ \ref{eq:Gss} must satisfy $(\partial_x-g)G = 0$, which is solved by
\begin{equation}
\label{eq:Gform}
G(x) = G(0) e^{gx},
\end{equation}
where $G(0)=e^{-g}$ is set by normalization:
\beq
G(0)e^g = G(1) = \sum_{n=0}^\infty p_n (1)^n = \sum_{n=0}^\infty p_n = 1.
\eeq
The steady state distribution is recovered via inverse transform:
\beq
p_n=\frac{1}{n!} \partial_x^n \left[ e^{g(x-1)} \right]_{x=0} = e^{-g}\frac{g^n}{n!},
\eeq
a Poisson distribution with parameter $g$, as before (Eqn.\ \ref{eq:pstar1}).

The time-dependent solution to Eqn.\ \ref{eq:Gss} may be obtained using the method of characteristics,
in which one looks for characteristic curves along which the partial differential equation
becomes an ordinary differential equation. The
curves are parameterized by $s$, such that
$G(x,t) = G(x(s),t(s)) = G(s)$, and we
look for the solution to the ordinary differential equation for $G(s)$, having eliminated $x$ and $t$ from the problem. First Eqn.\ \ref{eq:Gss} can be rewritten more explicitly as
\beq
\label{Gx2}
(x-1)gG = \frac{\partial G}{\partial t}+\frac{\partial G}{\partial x}(x-1).
\eeq
Using the chain rule on $G(s)$ gives
\beq
\label{chain}
\frac{dG}{ds} = \frac{\partial G}{\partial t}\frac{dt}{ds}+\frac{\partial G}{\partial x}\frac{dx}{ds}.
\eeq
Consistency of Eqns.\ \ref{Gx2} and \ref{chain} requires
\beqn
\label{ts}
\frac{dt}{ds}&=&1, \\
\label{xs}
\frac{dx}{ds}&=&x-1, \\
\label{Gs}
\frac{dG}{ds}&=&(x-1)gG.
\eeqn
Eqn.\ \ref{ts} implies $s=t$ (we set $t_0=0$ without loss of generality).  Therefore, defining $y\equiv x-1$, Eqn.\ \ref{xs} implies
\beq
\label{xs2}
y = y_0e^{t}.
\eeq
Finally,
straightforward integration of Eqn.\ \ref{Gs} yields
\beq
\label{Gtemp}
G = G_0\exp \left[g y_0 \int_{0}^t dt' \, e^{t'} \right]
= G_0\exp \left[g y_0 ( e^{t}-1) \right]
= F(y_0) \exp \left[ gy_0e^t \right].
\eeq
where
we define $F(y_0) \equiv G_0e^{-gy_0}$.   We may expand $F(y_0)$ as a function of its argument, i.e.\ $F(y_0) =\sum_{j=0}^\infty A_j y_0^j$ for some coefficients $A_j$, such that
\beq
G = \sum_j A_j y_0^j \exp \left[g y_0e^t \right].
\eeq
Inserting $y_0 = ye^{-t} = (x-1)e^{-t}$ (Eqn.\ \ref{xs2}), we obtain
\beq
\label{Gxeigen}
G(x,t) = \sum_j A_j e^{-jt} (x-1)^j e^{g(x-1)}.
\eeq 
For $t \rightarrow \infty$ only the $j=0$ term survives and we recover the steady state generating function (Eqn.\ \ref{eq:Gform}),
\beq
G(x) =c_0e^{g (x-1)}
\eeq
where $c_0 = 1$ by normalization.
Eqn.\ \ref{Gxeigen} constitutes the full time-dependent solution to the birth death process; the time-dependent distribution can be retrieved using the inverse transform, Eqn.\ \ref{invtrans}, and the coefficients  $A_j$ are computable from the initial distribution $p_n(0)$, which will be made explicit in the next section.  

Since Eqn.\ \ref{eq:Gss} is linear in $G$, one may arrive at its solution in a second, elegant way by expanding in the eigenfunctions of the $x$-dependent operator.
That is, writing Eqn.\ \ref{eq:Gss} as $\dot{G} = -\L G$, where
\begin{equation}
\label{eq:Lx}
{\cal L} = (x-1)(\partial_x-g),
\end{equation}
we expand $G$ in eigenfunctions $\phi_j(x)$ with time-dependent expansion coefficients $G_j(t)$,
\begin{equation}
\label{eq:Geig}
G(x,t) = \sum_j G_j(t) \phi_j(x),
\end{equation}
where the $\phi_j(x)$ satisfy
\begin{equation}
\label{eq:eigx}
{\cal L}\phi_j = \lambda_j\phi_j
\end{equation}
for eigenvalues $\lambda_j$.  Substituting Eqn.\ \ref{eq:Geig} into Eqn.\ \ref{eq:Gss}
gives
$\sum_j \dot{G}_j \phi_j = - \sum_j \lambda_j G_j \phi_j$,
which, by orthogonality of the $\phi_j$, yields the set of ordinary differential equations
$\dot{G}_j = -\lambda_j G_j$,
solved by
\begin{equation}
\label{eq:ft}
G_j(t) = A_j e^{-\lambda_j t}
\end{equation}
for some constants $A_j$.
Comparing the result,
\beq
G(x,t) = \sum_j A_j e^{-\lambda_j t} \phi_j(x),
\eeq
with Eqn.\ \ref{Gxeigen} reveals the forms of the eigenvalues
\beq
\label{eval1}
\lambda_j = j \in \{0,1,2,\dots\}
\eeq
and the eigenfunctions
\beq
\label{efunc1}
\phi_j(x) = (x-1)^je^{g(x-1)},
\eeq
facts that we will confirm in the next section using operator methods.

\subsubsection{Operator representation}\label{sec:oprep}

We now introduce an a representation of the master equation in terms of raising and lowering operators.  This representation makes the solution more elegant, yielding a simple algebra which allows calculation of projections between the spaces of protein numbers and eigenfunctions without explicit computation of overlap integrals; it also lays the formal groundwork for solving models of multidimensional regulatory networks.  The formalism was first used for diffusion by Doi \cite{doi} and Zel'dovich
\cite{Zeldovich} and later developed by Peliti \cite{peliti}.

As before the generating function is defined as an expansion in a complete basis indexed by protein number $n$ in which the expansion coefficients are the probabilities $p_n$.  Within the operator formalism, this basis is represented as the set of states $\ket{n}$, i.e.\
\begin{equation}
|G(t)\rangle = \sum_{n=0}^\infty p_n(t) |n\rangle.
\end{equation}
Here we have adopted state notation commonly used in quantum mechanics \cite{sakurai1985modern}; the previous representation (Eqn.\ \ref{eq:Gxdef}) is recovered by projecting onto this equation the conjugate states $\bra{x}$ with the provisions
\beqn
\label{eq:Gketdef}
\langle x|G(t)\rangle &\equiv& G(x,t), \\
\label{eq:xonn}
\langle x|n\rangle &\equiv& x^n.
\eeqn
In Appendix \ref{app:nbasis}
we show how the orthonormality of the $\ket{n}$ states,
\begin{equation}
\label{eq:ortho}
\langle n|n' \rangle 	= \delta_{nn'}, 
\end{equation}
dictates the form of the conjugate state:
\begin{equation}
\label{eq:nonx}
\langle n|x\rangle = \frac{1}{x^{n+1}}.
\end{equation}
Projecting $\bra{n}$ onto Eqn.\ \ref{eq:Gketdef} and using Eqn.\ \ref{eq:ortho} gives the inverse transform:
\begin{equation}
\label{eq:nGp}
\langle n|G(t)\rangle = p_n(t).
\end{equation}

The equation of motion in the operator representation is obtained by summing the master equation (Eqn.\ \ref{eq:master}) over $n$ against $\ket{n}$ (see Appendix \ref{app:summing}), giving
\beq
\label{eom_op}
\ket{\dot{G}} = -\Lop\ket{G},
\eeq
where
\begin{equation}
\label{eq:Lop}
\hat{\cal L} \equiv (\hat{a}^+ - 1)(\hat{a}^- - g).
\end{equation}
Just as in the operator treatment of the quantum harmonic oscillator \cite{sakurai1985modern}, the operators $\ad$ and $\al$ here raise and lower the protein number by 1 respectively, i.e.\
\begin{eqnarray}
\label{eq:a+right}
\hat{a}^+|n\rangle &=& |n+1\rangle, \\
\label{eq:a-right}
\hat{a}^-|n\rangle &=& n|n-1\rangle
\end{eqnarray}
(note, however, that the prefactors here, $1$ and $n$, are different than those conventionally used for the harmonic oscillator, $\sqrt{n+1}$ and $\sqrt{n}$, respectively).
Comparison of Eqns.\ \ref{eq:master_op3}, \ref{eq:Gss}, and \ref{eq:Lop} makes clear the following correspondences among the master equation, generating function, and operator representations respectively:
\begin{alignat}{2}
\label{correspond1}
E^- &\leftrightarrow\, x &\leftrightarrow\,\,& \ad, \\
\label{correspond2}
(n+1)E^+ &\leftrightarrow\, \partial_x &\,\,\leftrightarrow\,\,& \al.
\end{alignat}
While it might seem strange that the down-shift operator $E^-$ corresponds to the raising operator $\ad$ (and the up-shift operator to the lowering operator), in Appendix \ref{app:aproperties} we show that, just as with the quantum harmonic oscillator, $\ad$ and $\al$ lower and raise protein number respectively when operating to the left, i.e.\
\begin{eqnarray}
\label{eq:a+}
\langle n|\hat{a}^+ &=& \langle n-1|, \\
\label{eq:a-}
\langle n|\hat{a}^- &=& (n+1)\langle n+1|,
\end{eqnarray}
which makes the correspondence more directly apparent.
Finally, again as with the quantum harmonic oscillator, the raising and lowering operators enjoy the commutation relation (see Appendix \ref{app:aproperties})
\begin{equation}
\left[ \hat{a}^-, \hat{a}^+ \right] = 1,
\end{equation}
and $\ad\al$ acts as a number operator, i.e.\ $\ad\al\ket{n} = n\ket{n}$.

Eqn.\ \ref{eq:Lop} shows that the full operator $\Lop$ factorizes, suggesting the definition of the shifted
operators
\begin{eqnarray}
\label{eq:b+}
\hat{b}^+ &\equiv& \hat{a}^+ - 1, \\
\label{eq:b-}
\hat{b}^- &\equiv& \hat{a}^- - g.
\end{eqnarray}
Since $\hat{b}^+$ and $\hat{b}^-$ differ from $\hat{a}^+$ and $\hat{a}^-$, respectively, by scalars, they obey the same commutation relation, i.e.\
\begin{equation}
\label{bcom}
\left[ \hat{b}^-, \hat{b}^+ \right] = 1.
\end{equation}

Since the equation of motion (Eqn.\ \ref{eom_op}) is linear, it will benefit from expansion of $\ket{G}$ in the eigenfunctions $\ket{\lambda_j}$ of the operator $\Lop$, where
\begin{equation}
\label{eq:eigop}
\hat{\cal L}|\lambda_j\rangle = \lambda_j|\lambda_j\rangle
\end{equation}
for eigenvalues $\lambda_j$.
In Appendix \ref{app:bproperties} we show that the commutation relation (Eqn. \ref{bcom}) and the steady state solution $\ip{x}{G}=e^{g(x-1)}$, for which $\Lop\ket{G} = 0$, completely define the eigenfunctions and eigenvalues of $\hat{\L}$; we summarize the results of Appendix \ref{app:bproperties} here.  The eigenvalues are shown to be nonnegative integers $j$,
\beq
\label{eval2}
\lambda_j = j \in \{0,1,2,\dots\},
\eeq
as in Eqn.\ \ref{eval1}.  The eigenvalue equation now reads
\beq
\Lop\ket{j} = \bp\bl\ket{j} = j\ket{j},
\eeq
which is consistent with interpretation of $\bp\bl$ as a number operator for the eigenstates $\ket{j}$.
The eigenfunctions are shown to be (in $x$ space)
\beq
\label{efunc2}
\ip{x}{j} = (x-1)^je^{g(x-1)},
\eeq
as in Eqn.\ \ref{efunc1} with $\phi_j(x) \equiv \ip{x}{j}$.  The conjugate eigenfunctions are shown to be
\beq
\label{efunccon}
\ip{j}{x} = \frac{e^{-g(x-1)}}{(x-1)^{j+1}}.
\eeq
The operators $\bp$ and $\bl$ are shown to raise and lower the eigenstates $\ket{j}$, respectively, as $\ad$ and $\al$ do the $\ket{n}$ states, i.e.\
\beqn
\label{baction1}
\bp\ket{j} &=& \ket{j+1}, \\
\bl\ket{j} &=& j\ket{j-1}, \\
\bra{j}\bp &=& \bra{j-1}, \\
\label{baction4}
\bra{j}\bl &=& (j+1)\bra{j+1}.
\eeqn

Now employing the expansion of $\ket{G}$ in the eigenstates $\ket{j}$,
\beq
\label{Gexp_op}
\ket{G(t)} = \sum_j G_j(t)\ket{j},
\eeq
the equation of motion (Eqn.\ \ref{eom_op}) gives a trivial equation for the expansion coefficients
\beq
\label{Gjeig}
\dot{G}_j = -jG_j,
\eeq
which is solved by
\beq
G_j(t) = A_je^{-jt}
\eeq
for some constants $A_j$, as in Eqn.\ \ref{eq:ft}.  The inverse of Eqn.\ \ref{Gexp_op} (obtained by projecting $\bra{j}$ and using $\ip{j}{j'} = \delta_{jj'}$) is
\beq
\ip{j}{G(t)} = G_j(t).
\eeq

The probability distribution is retrieved by inverse transform (Eqn.\ \ref{eq:nGp}),
\beq
\label{pntfull}
p_n(t) = \ip{n}{G(t)} = \bra{n}\sum_j G_j(t) \ket{j} = \sum_j A_j e^{-jt} \ip{n}{j},
\eeq
where the coefficients $A_j$ are computed from the initial distribution $p_n(0)$:
\beq
\label{Ajcompute}
A_j = G_j(0) = \ip{j}{G(0)} = \sum_n p_n(0) \ip{j}{n}.
\eeq
Eqns.\ \ref{pntfull} and \ref{Ajcompute} give the full time-dependent solution to the birth--death process as expansions in the eigenmodes $\ip{n}{j}$ and conjugate eigenmodes $\ip{j}{n}$.  Because we have decomposed the problem using the eigenbasis, or spectrum, of the underlying operator, we will refer to this as the spectral solution.

The quantities $\ip{n}{j}$ and $\ip{j}{n}$ are overlaps between the protein number basis $\ket{n}$ and the eigenbasis $\ket{j}$.  Both $\ip{n}{j}$ and $\ip{j}{n}$ are readily computed by contour integration or more efficiently by recursive updating; these techniques are presented in Appendix \ref{app:overlap}.  Notable special cases are
\beqn
\label{special1}
\ip{n}{0} &=& e^{-g}\frac{g^n}{n!}, \\
\label{special2}
\ip{0}{n} &=& 1,
\eeqn
which confirm that Eqn.\ \ref{pntfull} describes a Poisson distribution in steady state ($j=0$).

\subsection{Fokker-Planck approximation}\label{FPapprox}
The previous sections discuss several methods for calculating the steady state and time dependent solutions of the master equation for the birth--death process. For many larger systems with regulation, full solution of the master equation is not possible. In some of these cases one can make progress by deriving an approximate equation which is valid when protein numbers are large.
In the limit of large protein numbers the master equation can be expanded to second order to yield the Fokker--Planck equation. In this section we derive the Fokker--Planck equation for a
general one-dimensional master equation
with arbitrary production and degradation rates;
the method is easily generalizable to models with regulation.

For arbitrary production rate $g_n$ and degradation rate $r_n$, the master equation reads
\beq
\partial_t p_n =-\left(g_n+r_n\right) p_n +g_{n-1} p_{n-1}+r_{n+1}p_{n+1};
\label{genbd}
\eeq
setting $g_n = g$ and $r_n = n$ recovers the simple birth--death process with time rescaled by the degradation rate, as in Eqn.\ \ref{eq:master}.

\subsubsection{Large protein number}\label{FPapprox1}

The Fokker--Plank equation is derived under the assumption that the typical protein number is large ($n \gg 1$), such that $n$ can be approximated as a continuous variable,
and a change of $1$ protein can be treated as a small change.
We will use parentheses when treating $n$ as continuous and a subscript when treating $n$ as discrete.
Under this approximation, the function $f(n\pm 1)$, where $f(n) \in \{ g(n)p(n), r(n)p(n) \}$, can be expanded to second order as
\beq
f(n\pm 1) = f(n) \pm \partial_n f(n) + \frac{1}{2}\partial_n^2 f(n).
\eeq
Inserting the results of the expansion into  Eqn.\ \ref{genbd}, we obtain
\beq
\partial_t  p(n) =-\partial_n \left[ v(n) p(n) \right]+\frac{1}{2}\partial^2_n \left[ D(n) p(n) \right],
\label{genbdfin}
\eeq
where $v(n) \equiv g(n)-r(n)$, an effective drift velocity, recovers the right-hand side of the deterministic equation,
and $D(n) \equiv g(n)+r(n)$, an effective diffusion constant, sets the scale of the fluctuations in protein number.
The drift term plays the role of an effective force. The diffusion term plays the role of an effective temperature: the larger it is, the larger the excursions a  single trajectory (of particle number versus time) takes from the mean.
Eqn.\ \ref{genbdfin} is the Fokker--Planck equation.

The steady state solution of Eqn.\ \ref{genbdfin} is easily obtained by noticing that the Fokker--Planck equation is a continuity equation of the form
\beq
\partial_t p= -\partial_n j,
\eeq
where $j(n)\equiv v(n)p(n)-\frac{1}{2}\partial_n \left[ D(n)p(n)\right]$ is the current of the probability.
In steady state ($\partial_t p = 0$) the current is constant, and since $p(n) \rightarrow 0$ and $\partial_n p(n) \rightarrow 0$ as $n \rightarrow \infty$ (typically more quickly than $v(n)$ or $D(n)$ diverges), the current vanishes for all $n$.  The steady state distribution is then found by direct integration, i.e.\
\beq
\label{FPss}
p(n) = \frac{N}{D(n)} \exp \left[2 \int_0^n dn' \frac{v(n')}{D(n')} \right],
\eeq
where $N$ is a normalization constant ensuring $\int_0^\infty dn\, p(n) = 1$.

In the simple birth--death process, for which $v(n) = g - n$ and $D(n) = g + n$, Eqn.\ \ref{FPss} evaluates to
\beq
\label{FPssbd}
p(n) = \frac{N}{g}\left( 1+\frac{n}{g}\right)^{4g-1} e^{-2n}.
\eeq
In Appendix \ref{app:gauss}, we show that the exact steady state (the Poisson distribution, Eqn.\ \ref{eq:pstar1}) and Eqn. \ref{FPssbd} have the same asymptotic limit for large protein number: a Gaussian distribution with mean and variance equal to $g$, the ratio of production to degradation rates.

\subsubsection{Small noise}\label{FPapprox2}

In addition to the approximation that the typical protein number is large, one may make the further approximation that the noise is small.  This is often referred to as the ``linear noise approximation'' 
\cite{vanKampen, elfehrenberg, Paulsson}
or ``small noise approximation''
\cite{tkacikwalczakbialek09, tkacikwalczakbialek10}.
Specifically, one assumes that the fluctuations $\eta$ in $n$ are small around the mean $\bar{n}$, i.e.
\beq
n=\bar{n}+\eta,
\eeq
where $\bar{n} \equiv \avg{n} = \sum_n n p_n$ (for brevity we use bar notation in this and several subsequent sections).
Since $dn/d\eta=1$ we have $p(n)=p(\eta)$, and the master equation (Eqn.\ \ref{genbdfin}) becomes an equation in $\eta$:
\beq
\partial_t  p(\eta) =-\partial_\eta \left[ v(\bar{n}+\eta) p(\eta) \right]
	+\frac{1}{2}\partial^2_\eta \left[ D(\bar{n}+\eta) p(\eta) \right],
\label{master_eta}
\eeq
We use the approximation that $\eta$ is small to expand the drift and diffusion terms to first nonzero order in $\eta$:
\beqn
\label{vbar}
v(\bar{n}+\eta) &=& v(\bar{n}) + \eta v'(\bar{n}) + \dots \approx \eta v'(\bar{n}) \\
D(\bar{n}+\eta) &=& D(\bar{n}) + \eta D'(\bar{n}) + \dots \approx D(\bar{n})
\eeqn
where prime denotes differentiation with respect to $n$, and the last step in Eqn.\ \ref{vbar} recalls the fact that $v(\bar{n}) = g(\bar{n}) - r(\bar{n}) = 0$ in steady state, as given by the kinetic rate equation (e.g.\ Eqn.\ \ref{avgn}).  Eqn.\ \ref{master_eta} is now
\beq
\partial_t  p(\eta) =-v'(\bar{n})\partial_\eta \left[ \eta p(\eta) \right]
	+\frac{1}{2} D(\bar{n}) \partial^2_\eta \left[ p(\eta) \right].
\label{master_eta2}
\eeq
As with Eqn.\ \ref{FPss}, the steady state of Eqn.\ \ref{master_eta2} is found by direct integration:
\beq
\label{LNA}
p(\eta) = \frac{N}{D(\bar{n})} \exp \left[\frac{2v'(\bar{n})}{D(\bar{n})} \int_0^\eta d\eta' \eta' \right]
	= \frac{1}{\sqrt{2\pi \sigma^2}}e^{-\eta^2/2\sigma^2},
\eeq
where
\beq
\label{LNAwidth}
\sigma^2 \equiv \frac{D(\bar{n})}{-2v'(\bar{n})}
\eeq
 (note that $v'(\bar{n})$ is negative for stable fixed points).  Eqn.\ \ref{LNA} is a Gaussian distribution, meaning that under the linear noise approximation, the steady state probability distribution is a Gaussian centered at the exact mean $\bar{n}$ and with width determined by mean birth and death rates according to Eqn. \ref{LNAwidth}.  We note that Eqn.\ \ref{LNA} can be equivalently derived by expanding the integrand in Eqn.\ \ref{FPss} about its maximum (or ``saddle point'') $\bar{n}$.

The linear noise approximation is stricter than the large-protein number approximation made in the previous section (Sec.\ \ref{FPapprox1}). While the previous approximation makes no assumption about the form of the probability distribution, the linear noise approximation assumes that the distribution is unimodal and sharply peaked around its mean. In practice,
$\bar{n}$ and $\sigma^2$ are obtained by finding the steady state(s) (i.e. the stable fixed point(s)) of
the corresponding deterministic rate equation.
However, it is easily possible (for processes more complicated than the simple birth--death process) for the deterministic equation to have more than one stable fixed point (see Fig.\ \ref{bif}). Although one may make a Gaussian expansion around each fixed point in turn, the linear noise approximation does not describe how these Gaussians might be weighted in a multi-modal distribution. In these cases it is most accurate to use (if solvable) either the large-protein number approximation (Eqn.\ \ref{FPss}) or the original master equation (Eqn.\ \ref{genbd}).

In the simple birth--death process, for which $v(n) = g - n$, $D(n) = g + n$, and $\bar{n} = g$, Eqn.\ \ref{LNAwidth} gives $\sigma^2 = 2g/[-2(-1)] = g$, and therefore Eqn.\ \ref{LNA} reproduces the asymptotic behavior derived in Appendix \ref{app:gauss}.

\subsection{Langevin approximation}\label{sec:Langevin}
We now consider a second stochastic approximation: the Langevin equation. The advantage of the Langevin approach is that a large amount can be learned about the process without finding the full distribution, but instead by considering the correlation functions of the concentration, which are readily computed. 
 Starting from the Fokker--Planck equation we can calculate the equation for the mean of the distribution $\avg{n}=\sum_{n=0}^{\infty} n p_n$. We arrive at the kinetic rate equation given in Eqn.\ \ref{kinone}. 

We can think about the change in concentration in time as a trajectory in $n$ space. Each realization of a birth-death process, will be described by a certain $n(t)$, and averaging over many experiments, we obtain $\avg{n(t)}$, as given by Eqn.\ \ref{kinone}. If we consider the change of concentrations on time scales longer than the characteristic timescales of the particular reactions, we can assume that there are many birth and death processes in each interval and that the fluctuations of each realization of $n(t)$ around the mean are Gaussian distributed:
\beq
\label{trajnoise}
P\left[\eta\right]\sim e^{-\int dn \int dt \frac{\eta^2(n, t)}{4D(n)}}
\eeq
We can include fluctuations around the the mean values by considering an additional noise term $\eta(t)$, such that the equation for the change of the concentration of proteins $n$ becomes:
\beq
\frac{d n}{dt}=v(n)+\eta(t)=g-n+\eta(t).
\label{Langevin}
\eeq
We require of the noise that:
\begin{eqnarray}
\label{defLang}
\avg{\eta(t)}&=&0, \\
\avg{\eta(t') \eta(t)}&=&\delta(t-t') D(n)=\delta(t-t') (g+n)
\end{eqnarray}
where $D(n)$ is in general the diffusion term in the Fokker--Planck equation. In Appendix \ref{Appendix_LangtoFP} we show the equivalence of the Fokker--Planck and Langevin descriptions.

In the general case when there are many types of proteins in the system, we can define a time dependent correlation function
\beq
C_{ij}(t)=\avg{\delta N_i(0)\delta N_j(t)},
\eeq
where the average implies a time average and $\delta N_i(t)=  N_i(t)-\avg{\delta N_i(t)}$ are the variances of each type of protein. In one dimension $\delta n^2$ is the variance $\sigma^2$ (in subsequent sections we use $\delta n^2$ and $\sigma^2$ interchangeably).  We note that in steady state the time average can be replaced by an ensemble average. For the case of the single species birth--death process we have already computed the means and variances in Sec.\ \ref{steadystateBD}. In case of the birth--death process, we calculate the time dependent correlation function to obey the equation:
\beq
\label{autocor}
\frac{d C(t)}{dt}=-C(t).
\eeq
In steady state the solution must reduce to the previously calculated variance (Appendix \ref{Appendix_meansofBD}): $\delta n^2=\avg{(n-\avg{n})((n-\avg{n}))}= \avg{n^2}-\avg{n}^2=g$. Therefore the solution of Eqn.\ \ref{autocor} for a simple birth--death process is
\beq
C(t)=ge^{-t}.
\eeq

We can also consider the correlation functions in Fourier space 
\beq
\label{corrres}
C_{ij}(\omega)=\int_0^{\infty} dt e^{-i\omega t} C_{ij}(t).
\eeq
Often it is easier to calculate the Fourier transform of the correlation function directly from the Fourier transform of the Langevin equation, and then invert the transform back. In case of the simple birth--death process we can vary the Langevin equation around its mean values $ n(t)= \avg{n}+\delta n(t)$ (that is simply linearize the equation around its mean) and consider the resulting equation in Fourier space to obtain
\beq
-i \omega \delta \tilde{n} (\omega)=- \delta \tilde{n} (\omega)+\tilde{\eta}(\omega),
\eeq
where $ \delta  \tilde{n} (\omega)= \int_0^{\infty} dt e^{-i\omega t} n(t)$ and 
\beqn
\tilde{\eta}(\omega)&=& \int_0^{\infty} dt e^{-i\omega t} \eta (t)\\
\avg{\tilde{\eta}(\omega) \tilde{\eta}(\omega')}&=&2 \pi \delta(\omega-\omega') (g+ \avg{n}).
\eeqn
The correlation function is then:
\beq
\avg{ \delta  \tilde{n}^* (\omega)   \delta \tilde{n} (\omega')}=2 \pi \delta(\omega- \omega')\frac{g+ \avg{n}}{\omega^2+1 }=2 \pi \frac{2g}{\omega^2+1 }.
\eeq
Inverting the Fourier transform reproduces the real time correlation function.
\beq
C(t)=\int_0^{\infty} d\omega e^{i\omega t} \frac{2g}{\omega^2+1 }=2 \pi i e^{-t} \frac{2g}{2i}=g e^{-t},
\eeq
where we reproduce the result of Eqn.\ref{corrres}.
The real part of the auto correlation function in Fourier space is also called the power spectrum
\beq
{\cal{N}}(\omega)= \avg {\delta \tilde{n}^* (\omega) \delta  \tilde{n} (\omega')}= \int_0^{\infty} dt e^{-i\omega t} C_{ii} (t),
\eeq
because it tells us which frequency modes contribute most to the form of the noise. We note that the power spectrum $\cal{N}(\omega)$ may be written as integral of the correlation function in real time.

\subsection{Comparison of the descriptions}
\begin{figure}
\includegraphics[scale=1]{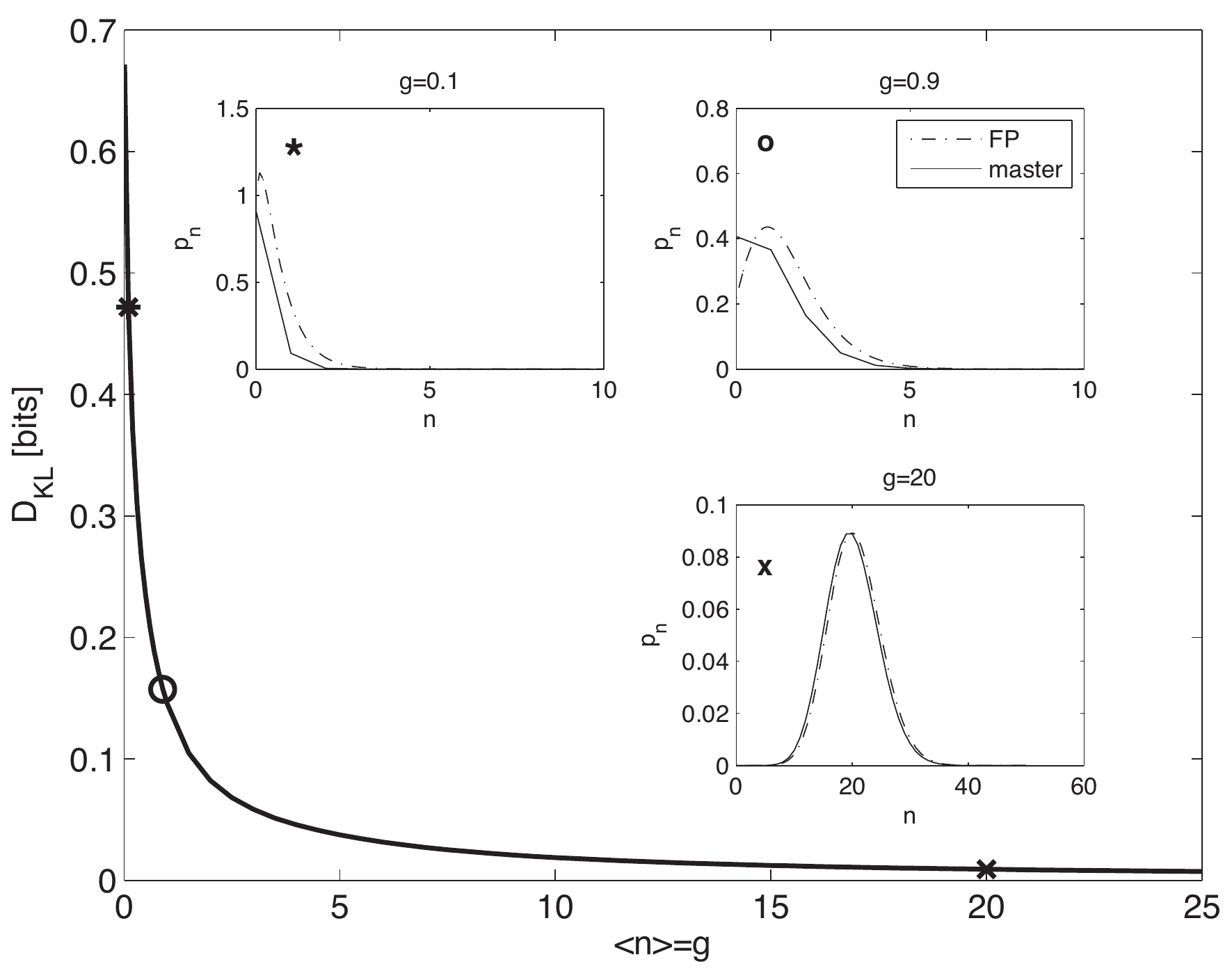}
\caption{Comparison of the distributions obtained from the master equation (Eqn.\ \ref{eq:pstar1}) and from the Fokker--Planck approximation (Eqn.\ \ref{FPss}),
for the simple birth--death process. The Kullback--Leibler divergence $D_{KL}$ (Eqn.\ \ref{DKL}) between 
the two distributions is plotted as a function of mean protein number $\avg{n} = g$.
Insets show distributions obtained from the master equation (solid) and from the Fokker--Planck equation (dashed) for
$g=0.1$ (star), $g=0.9$ (circle), and $g=20$ (cross), where the symbols correspond to points on the $D_{KL}$ curve.}
\label{FigBD}
\end{figure}

For the birth--death process, the steady state of the kinetic rate equation agrees with the mean of the steady state probability distribution (under both the exact and approximate stochastic descriptions):
$\avg{n}=g$.
In the limit of large protein number,
the Poisson distribution is well approximated by a Gaussian, and the Fokker--Planck (and therefore the Langevin) approximation is a good description. To investigate the validity of the Fokker--Planck approximation at small and intermediate protein number, in Fig.\ \ref{FigBD} we compare for a range of protein numbers the probability distribution obtained directly from the master equation (i.e.\ the Poisson distribution, Eqn.\ \ref{eq:pstar1}) and from the Fokker--Planck approximation (Eqn.\ \ref{FPss}). We quantify the disagreement between two distributions using the Kullback--Leibler divergence,
\beq
\label{DKL}
D_{KL}=\sum_n p_n \log \frac{p_n}{\tilde{p}_n},
\eeq
where $p_n$ corresponds to the master equation and $\tilde{p}_n$ corresponds to the Fokker-Planck approximation.
The Kullback--Leibler divergence is not symmetric and is therefore appropriate for a comparison between a ``true'' distribution ($p_n$ here) and its approximation ($\tilde{p}_n$ here).
As the mean protein number $\avg{n}=g$ increases, the accuracy of the approximation increases, and the divergence $D_{KL}$ decreases. We plot explicitly three sample pairs of exact and approximate distributions, at small, intermediate, and large protein numbers.
As expected the Fokker-Planck distribution deviates from the Poisson distribution at small protein number, and agrees well at large protein number.

\section{Autoregulation}\label{BDreg}
We now begin to turn our attention from the simple one-dimensional birth--death process to more realistic models gene regulation (see \cite{BintuHwa1, BintuHwa2} for a discussion of regulation functions of gene expression). We start with the description of auto-regulation of one gene, in which its protein production rate is an arbitrary function of its own protein number.

\subsection{Deterministic model}

As before the mean dynamics are captured by the kinetic rate equation.  The kinetic rate equation for a birth--death process in which the production rate is an arbitrary function $g(\bar{n})$ of mean protein number $\bar{n}$ is
\beq
\frac{d\bar{n}}{dt}=g(\bar{n})-\bar{n}.
\label{kintwo}
\eeq
The problem becomes potentially much harder, since the auto-regulation function $g(\bar{n})$ can be nonlinear.

The form of the auto-regulation function depends specifically on the molecular model of regulation which is being considered.   For example, the Hill function,
\beq
\label{regfun}
g(n)= \frac{g_-K^h+g_+ n^h}{K^h+n^h},
\eeq
can be derived by considering a gene with two production rates $g_+$ and $g_-$ (corresponding to the states in which a protein is bound and unbound to the DNA, respectively)
for which the rate of switching to the bound state depends on the protein number (see Sec.\ \ref{bursts} and Appendix \ref{Hillderiv}).  The derivation assumes that the rates of binding and unbinding are faster than the protein degradation rate (which sets the time-scale for changes in protein number), such that equilibrium is reached, with equilibrium binding constant $K$. 

The parameter $h$ describes the cooperativity of protein binding.
When $g_+ > g_-$,
the case $h>0$ corresponds to activation and the case $h<0$ corresponds to repression; the special case $h=0$ reproduces the simple birth--death process.  For $|h| \ge 2$, Eqn.\ \ref{kintwo} has two stable fixed points for certain parameter regimes; for $|h| \ge 2$ fixed points must be found numerically.

Although the Hill equation is often used in models of gene regulation, we note that other functional forms are derivable from other biochemical processes.  Many of the following results are valid for arbitrary $g(n)$.

\subsection{The master equation}\label{automaster}
The full stochastic description corresponding to the deterministic Eqn.\ \ref{kintwo} is given by the birth--death master equation in which the production rate is described by the arbitrary auto-regulation function $g_n$ (recall that we replace parentheses with subscript when treating $n$ as discrete):
\begin{equation}
\label{eq:masterself}
\dot{p}_n = -g_n p_n - np_n +g_{n-1} p_{n-1} + (n+1)p_{n+1},
\end{equation}
We may easily generalize the solution in Sec.\ \ref{steadystateBD} to find the steady state probability distribution,
\beq
\label{pn}
p_n = \frac{p_0}{n!}\prod_{n'=0}^{n-1}g_{n'},
\eeq
with $p_0$ set by normalization. Except for special cases of the regulation function we cannot find a closed form solution for the distribution, but the product is easily evaluated.

The results in Sec.\ \ref{BD} for the simple birth--death process
can be generalized in the case of auto-regulation.
We will pay particular attention to the generalization of the eigenfunction expansion in operator notation for arbitrary $g_n$, when discussing two genes in Sec.\ \ref{twogenes}.

\subsection{Bistability and noise}

Auto-regulation can affect the statistics of the steady state distribution.  Even before specifying the form of $g(n)$, we may obtain a general statistical result in the limit of large protein number by using the linear noise approximation (Sec.\ \ref{FPapprox2}).  Eqn.\ \ref{LNAwidth} describes the variance $\sigma^2$ of fluctuations around the steady state mean $\bar{n}$.  Here we compute the ratio $\sigma^2/\bar{n}$ for auto-regulation in order to compare with the Poisson distribution, for which $\sigma^2/\bar{n} = 1$.

\begin{figure}
\includegraphics[scale=1]{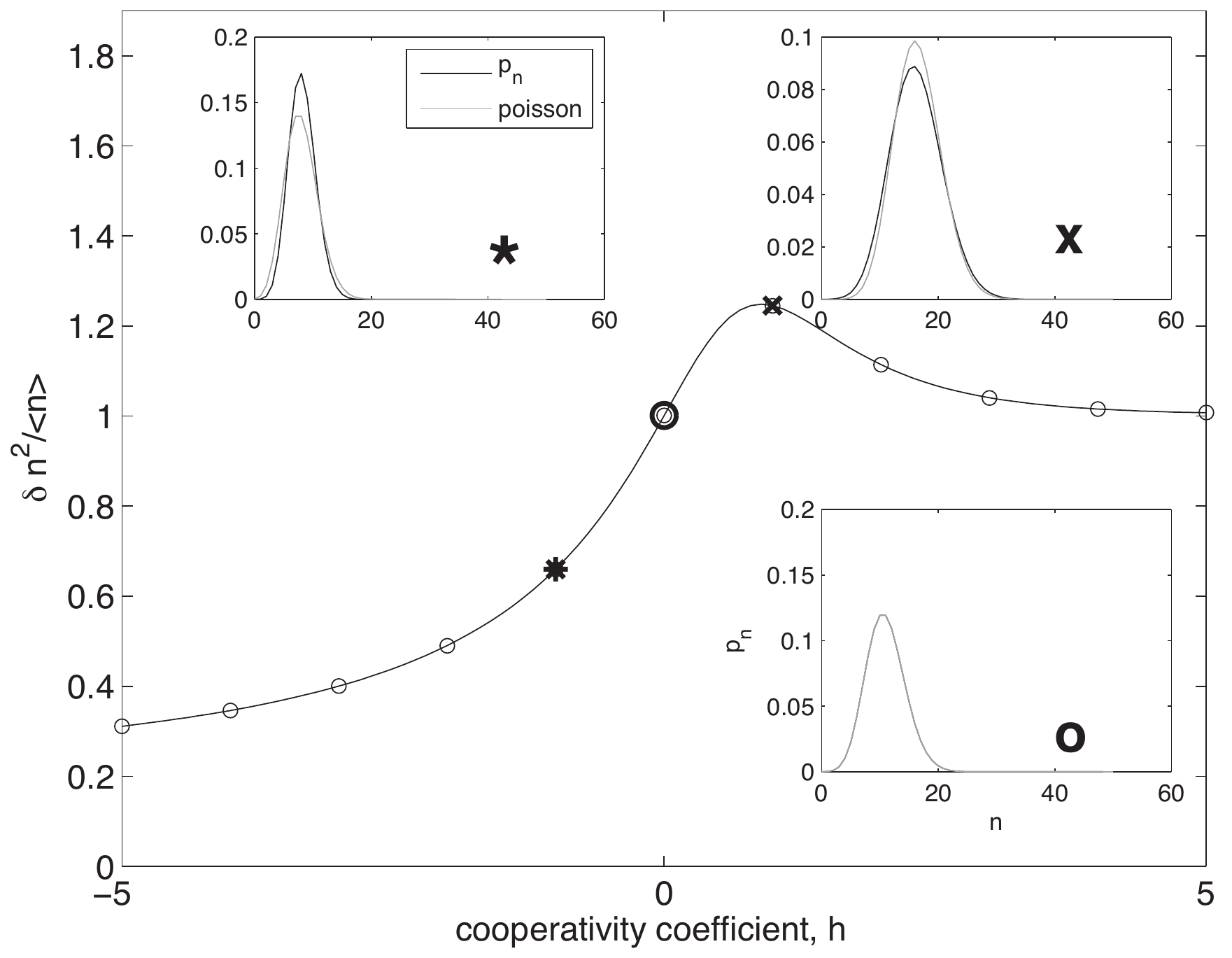}
\caption{The ratio of the variance $\delta n^2$ to the mean $\avg{n}$ as a function of the cooperativity coefficient $h$, for a gene undergoing Hill function (Eqn.\ \ref{regfun}) autoregulation. For $h<0$ the gene represses its own production; for $h>0$ the gene activates its own production.
The case $h=0$ recovers the simple birth-death process result $\delta n^2/\avg{n}=1$. Points are plotted for integer values of $h$; the line shows Eqn.\ \ref{regfun} for continuous $h$. Insets show probability distributions (black solid lines) from the master equation, compared with Poisson distributions with the same mean (gray solid lines), at $h=-1$ (star), $h=0$ (circle) and $h=1$ (cross), where symbols correspond to points on the $\delta n^2/\avg{n}$ curve.
Parameters are $g_-=2$, $g_+=20$, and $K=4$, which restrict solutions to the mono-stable regime, yielding unimodal distributions.}
\label{meanvar}
\end{figure}

For auto-regulation the diffusion term becomes $D(\bar{n}) = g(\bar{n})+\bar{n} = 2\bar{n}$, where we have used the fact that $0 = g(\bar{n})-\bar{n}$ at steady state.  The derivative of the drift term evaluated at the mean is $v'(\bar{n}) = g'(\bar{n}) - 1$, where the prime denotes differentiation with respect to $n$.  Thus Eqn.\ \ref{LNAwidth} becomes
\beq
\label{supersub}
\frac{\sigma^2}{\bar{n}} = \frac{1}{1-g'(\bar{n})}.
\eeq
This expression shows that self-activation ($g'(\bar{n})>0$) leads to super-Poissonian noise ($\sigma^2/\bar{n}>1$), and self-repression ($g'(\bar{n})<0$) leads to sub-Poissonian noise ($\sigma^2/\bar{n}<1$); when there is no regulation we have $g'(\bar{n})=0$, and we recover the Poisson result, $\sigma^2/\bar{n}=1$. Fig.\ \ref{meanvar} demonstrates this behavior by computing the exact distribution (Eqn.\ \ref{pn}) for Hill function auto-regulation (Eqn.\ \ref{regfun}) with various values of the cooperativity parameter $h$.

\begin{figure}
\includegraphics[scale=1]{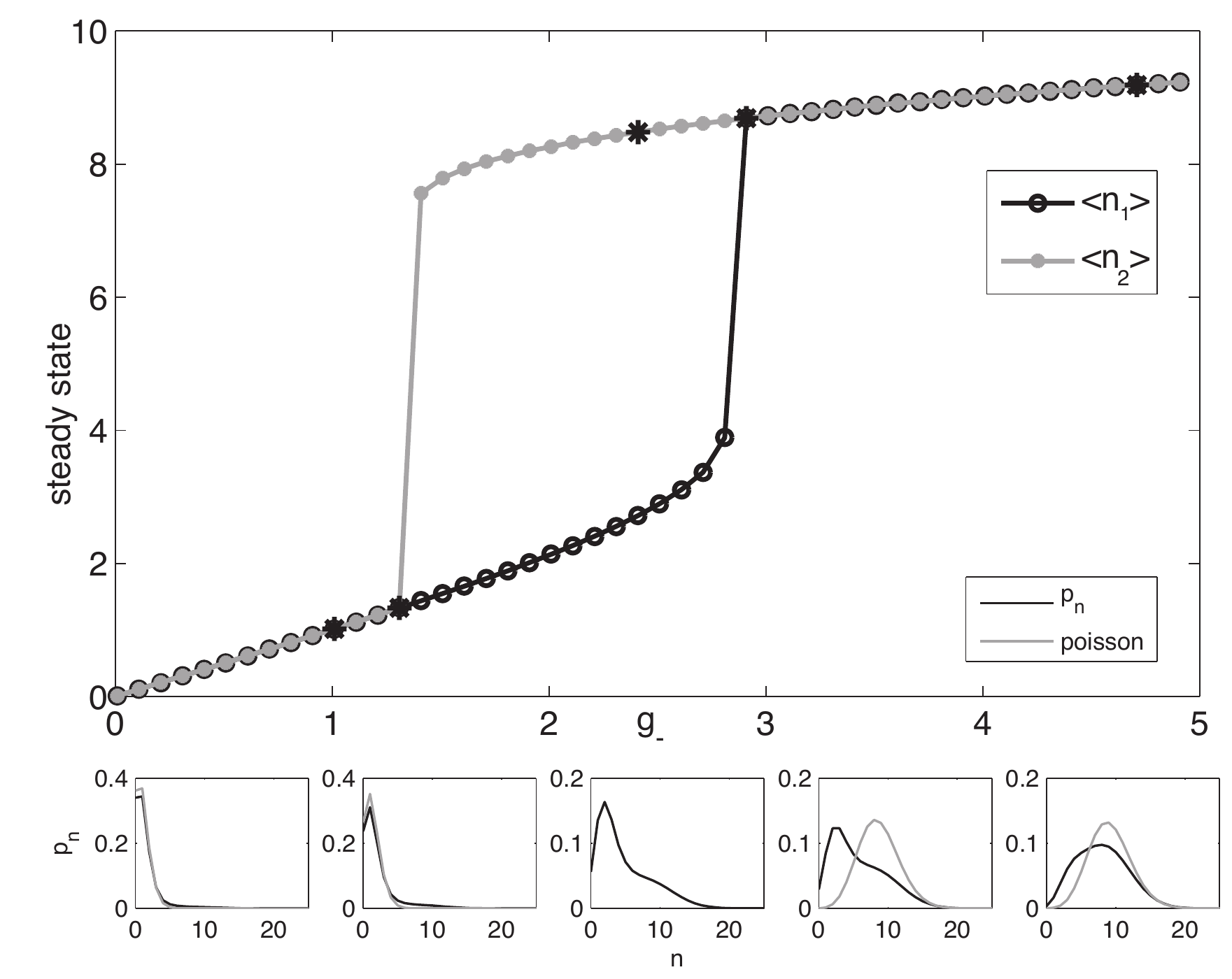}
\caption{The steady state solution(s) for a self-activating gene as a function of the basal rate $g_-$. The figure shows regimes which are mono-stable (high and low $g_-$) and bistable (intermediate $g_-$). The probability distributions below (solid black lines) correspond consecutively to the points above marked by stars.
In the mono-stable regime the distributions are compared to Poisson distributions (solid gray lines). The auto-regulation function is a Hill function (Eqn.\ \ref{regfun}) with $h=4$, $g_+=10$, and $K=6$.}
\label{bif}
\end{figure}

Eqn.\ \ref{supersub} diverges when $g'(\bar{n})=1$. This corresponds to a bifurcation point for the self-activating gene, where there are two possible solutions to the steady state equation. In this case the ratio of the variance to the mean is not very informative about the distribution because the distribution is bimodal. The self-repressing gene, in contrast, cannot be bistable, and the ratio of the variance to the mean is always a good description of noise. In Fig.\ \ref{bif} we demonstrate that the exact distribution becomes bimodal when the deterministic equation crosses a bifurcation point.

Eqn.\ \ref{supersub} can be be equivalently obtained by computing the power spectrum from the linearized Langevin equation.  Linearizing the Langevin equation and going to Fourier space we obtain
\beq
-i \omega \delta \tilde{n} (\omega)= g'(\bar{n})  \tilde{n} (\omega)-  \tilde{n} (\omega) +\tilde{\eta}(\omega).
\eeq
The power spectrum becomes
\beq
{\cal{N}}=\frac{2\bar{n}}{\omega^2+[1-g'(\bar{n})]^2}
\eeq
The steady state auto--correlation function is therefore
\beq
\delta n^2=\frac{\bar{n}}{\abs{1-g'(\bar{n})}},
\eeq
and the ratio of the variance to the mean is
\beq
\frac{\delta n^2}{\bar{n}}=\frac{1}{\abs{1-g'(\bar{n})}}.
\eeq
We point the reader to the paper of Warren, Tanase-Nicola and ten Wolde \cite{warrensorintenwolde06} for a pedagogical discussion of power spectra in gene expression.

\section{Bursts}\label{bursts}

The previous section assumes that the processes of transcriptional and translational regulation can be described by one deterministic regulation function.  Recent experiments have observed a feature of gene regulation that is not captured by such a model: protein production often occurs in bursts \cite{Raj, raj2, Golding}.  Bursts can result from the gene transitioning among two or more states with different rates of protein production, for example when a transcription factor is bound or unbound, or, in higher eukaryotes, when different production states are introduced by chromatin remodeling \cite{raj2}.  Bursts can also arise during translation, in which a single mRNA transcript produces many copies of the protein in a short amount of time.  Here we present simple models of the first type, which we call transcriptional bursts, and the second type, which we call translational bursts.  We then discuss a model which combines bursting with auto-regulation.
We refer the reader to the literature for discussion of other burst models
including those which capture both transcriptional and translational bursts
\cite{walczak_bj, mehta08}.

\subsection{Transcriptional bursts}

In this section we consider a gene that can exist in multiple states, each with its own protein production rate.  We first specialize to the case of two states, which is commonly used as a model for bursting \cite{Raj,IyerBiswas}.  We solve this case conventionally using a generating function.  The conventional solution is limited to two states; for an arbitrary number of states we present the spectral solution using operator methods.  Finally we present a solution to a simple model with both bursting and auto-regulation.

\subsubsection{The master equation}
We begin by considering the general process in which a gene has $Z$ different states. In each state $z$ the protein production is described by a simple birth-death process (Sec.\ \ref{BD}) with production rate $g_z$. Transitions between the states occur at rates given by the transition matrix $\Omega_{zz'}$, such that
\beq
\sum_{z=1}^Z \Omega_{zz'}=0,
\eeq
which follows from conservation of probability. The rates are constant here; in Sec.\ \ref{burstauto} we extend the model such that the rates depend on protein number.

The system is described by a $Z$-state probability vector with elements given by ${p}_n^z$. The particular states evolve according to the coupled master equation
\beq
\label{burstME}
\dot{p}_n^z = -g_zp_n^z - np_n^z + g_zp_{n-1}^z + (n+1)p_{n+1}^z + \sum_{z'}\Omega_{zz'} p_n^{z'}.
\eeq
where all but the last term describes the birth--death process for state $z$, and the last term describes transitions among states.
The probabilities of being in state $z$ regardless of the number of proteins are given by $\pi_z = \sum_np_n^z$; summing the master equation over $n$ in steady state gives
\beq
\label{pi1}
\sum_{z'}\Omega_{zz'}\pi_{z'}=0,
\eeq
and normalization requires
\beq
\label{pi2}
\sum_z\pi_z = 1.
\eeq

\subsubsection{The two-state gene}\label{on/off}
We first consider the case in which the gene has only two states ($Z=2$), an active state ($z = +$) and an inactive state ($z = -$):
\beq
\label{ppm}
p^z_n = \begin{pmatrix} p_n^+ \\ p_n^- \end{pmatrix}.
\eeq
The transition matrix takes the form
\beq
\label{Omega2}
\Omega_{zz'} = \begin{pmatrix} -\omega_- & \omega_+ \\ \omega_- & -\omega_+ \end{pmatrix}.
\eeq
where $\omega_+$ and $\omega_-$ are the transition rates to the active and to the inactive states, respectively.

As in Sec.\ \ref{sec:Gx}, we define the generating function $G_\pm(x) = \sum_n p_n^\pm x^n$ and sum the master equation (Eqn.\ \ref{burstME}) over $x^n$, yielding at steady state
\beq
\label{G2state}
0 = -y(\partial_y - g_\pm) G_\pm \pm \omega_+ G_- \mp \omega_- G_+,
\eeq
where $y \equiv x-1$.  Writing $G_\pm(y) = e^{g_\pm y} H_\pm(y)$ yields a simpler equation for $H_\pm(y)$; writing out the $+$ and $-$ cases explicitly,
\beqn
\label{H1}
0 &=& -y e^{y \Delta} \partial_y H_+ + \omega_+ H_- - \omega_- e^{y \Delta} H_+, \\
\label{H2}
0 &=& -y e^{-y \Delta} \partial_y H_- + \omega_- H_+ - \omega_+ e^{-y \Delta} H_-,
\eeqn
where $\Delta \equiv g_+ - g_-$.  Eqns.\ \ref{H1}-\ref{H2} are a coupled pair of first-order ordinary differential equations; combining them yields a second-order differential equation.  Specifically, solving Eqn.\ \ref{H1} (\ref{H2}) for $H_-$ ($H_+$) and substituting it into Eqn.\ \ref{H2} (\ref{H1}), one obtains
\beq
\label{hyper}
0 = u \partial_u^2 H_\pm + (\beta - u) \partial_u H_\pm - \alpha H_\pm,
\eeq
where $u \equiv \mp y \Delta$, $\alpha \equiv \omega_\mp$, and $\beta \equiv \omega_+ + \omega_- + 1$.
Eqn.\ \ref{hyper} is the canonical equation for the confluent hypergeometric function.  Only one of its two solutions satisfies $p_n \rightarrow 0$ as $n \rightarrow \infty$; it is $H_\pm(x) = N_\pm\Phi[\alpha,\beta;u]$, where
\beq
\label{hyp}
\Phi[\alpha,\beta;u] = \sum_{\l=0}^\infty \frac{\Gamma(\l+\alpha)}{\Gamma(\alpha)}\frac{\Gamma(\beta)}{\Gamma(\l+\beta)}\frac{u^\l}{\l!}
\eeq
is the confluent hypergeometric function of the first kind, and the scalars $N_\pm$ are found by normalization: $N_\pm = G_\pm(1) = \sum_n p_n^\pm (1)^n = \pi^\pm = \omega_\pm / (\omega_+ + \omega_-)$ (the last step uses Eqns.\ \ref{pi1}-\ref{pi2}).  Thus,
\beq
\label{Gpm}
G_\pm(x) = \frac{\omega_\pm}{\omega_+ + \omega_-} e^{g_\pm (x-1)}
	\Phi[\omega_\mp, \omega_+ + \omega_- + 1; \mp(g_+ - g_-)(x-1)].
\eeq
As before, the probability distribution can be recovered from $G(x) = \sum_\pm G_\pm(x)$ by differentiation (Eqn. \ref{invtrans}).

As shown in Appendix \ref{app:sri}, in the limit $g_- = 0$, Eqn.\ \ref{Gpm} reduces to the slightly simpler expression
\beq
\label{sri}
G(x) = \Phi[\omega_+,\omega_++\omega_-;g_+(x-1)],
\eeq
for which the probability distribution has the explicit form
\beqn
\label{raj}
p_n &=& \frac{g_+^n}{n!}\frac{\Gamma(n+\omega_+)}{\Gamma(\omega_+)}\frac{\Gamma(\omega_++\omega_-)}{\Gamma(n+\omega_++\omega_-)}
\Phi[\omega_++n,\omega_++\omega_-+n;-g_+],
\eeqn
in agreement with the results of Iyer-Biswas et al.\ \cite{IyerBiswas} and Raj et al.\ \cite{Raj}.

\subsubsection{The spectral solution}
It is clear that the method presented in the previous section results in a $Z$th order ordinary differential equation, and therefore its utility is generally limited to $Z=2$ states.  In contrast, here we show that the spectral solution, in which we expand in eigenfunctions of the birth--death operator, can be used for an arbitrary number of states.

We begin as in Sec.\ \ref{sec:oprep} by defining $\ket{G_z} = \sum_n p_n^z \ket{n}$ and writing the master equation in terms of raising and lowering operators,
\beq
\label{Gzop2}
\ket{\dot{G}_z} = -\bp \bl_z\ket{G_z}+\sum_{z'}\Omega_{zz'}\ket{G_{z'}},
\eeq
where now $\bl_z = \al - g_z$ is $z$-dependent.
We expand the generating function in the $z$-dependent eigenfunctions of the birth--death process as
\beq
\label{expandz}
\ket{G_z} = \sum_jG_j^z\ket{j_z},
\eeq
where the eigenvalue relation reads
\beq
\label{eigen}
\bp\bl_z\ket{j_z} = j\ket{j_z}.
\eeq
This gives the following equation for the expansion coefficients:
\beq
\label{cEoM}
\dot{G}_j^z = -jG_j^z + \sum_{z'}\Omega_{zz'}\sum_{j'}G_{j'}^{z'} \ip{j_z}{j'_{z'}}.
\eeq
Here the transition matrix $\Omega_{zz'}$ couples the otherwise independent birth--death processes, and
\beq
\label{overlap}
\ip{j_z}{j'_{z'}} = \frac{(-\Delta_{zz'})^{j-j'}}{(j-j')!}\theta(j-j'),
\eeq
where $\Delta_{zz'} = g_z - g_{z'}$, and the convention $\theta(0) = 1$ is used for the Heaviside function.  Eqn.\ \ref{overlap} is derived by evaluating the inner product using the $x$ space representations of the eigenfunctions,
\beqn
\label{jket}
\ip{x}{j_z} &=& (x-1)^j e^{g_z(x-1)},\\
\label{jbra}
\ip{j_z}{x} &=& \frac{e^{-g_z(x-1)}}{(x-1)^{j+1}},
\eeqn
and Cauchy's theorem, as in Appendix \ref{app:nbasis}.

The Heaviside function makes Eqn.\ \ref{cEoM} lower-triangular.  This is explicitly clear, for example, in steady state,
\beq
\label{css}
jG_j^z - \sum_{z'}\Omega_{zz'}G_j^{z'} = \sum_{z'\ne z}\Omega_{zz'}\sum_{j'<j}G_{j'}^{z'}\frac{(-\Delta_{zz'})^{j-j'}}{(j-j')!},
\eeq
where one observes that the $j$th component is a function only of components $j' < j$.  The lower-triangular structure allows $G_j^z$ to be computed iteratively, which makes the spectral solution numerically efficient.  The lower-triangular structure is a consequence of rotating to the eigenbasis of the birth-death operator; this structure was not present in the original master equation.  Other spectral decompositions are possible by exploiting other eigenbases; for example by expanding in a single eigenbasis parameterized by a constant (not $z$-dependent) production rate, one obtains an equation that is sub-diagonal in $j$ \cite{cspan}.

The probability distribution is recovered via inverse transform,
\beq
p_n^z = \sum_jG_j^z\ip{n}{j_z},
\eeq
as in Sec.\ \ref{sec:oprep}.
The spectral solution is valid for any number of states $Z$, while direct solution of the differential equation is generally limited to $Z=2$.
As expected, for $Z=2$ the spectral solution recovers the confluent hypergeometric function solution (Eqn.\ \ref{Gpm}), as shown in Appendix \ref{app:sri}.

Fig.\ \ref{fig:3state} demonstrates the spectral solution for $Z=3$ states when the transition rate among states is (a) equal to, and (b) much less than, the degradation rate.  In (a) the bursts give rise to a long tail in the distribution (compared to the Poisson distribution); in (b) the system spends long times in each production state before transitioning, resulting in a trimodal distribution.

\begin{figure}
\includegraphics[scale=1]{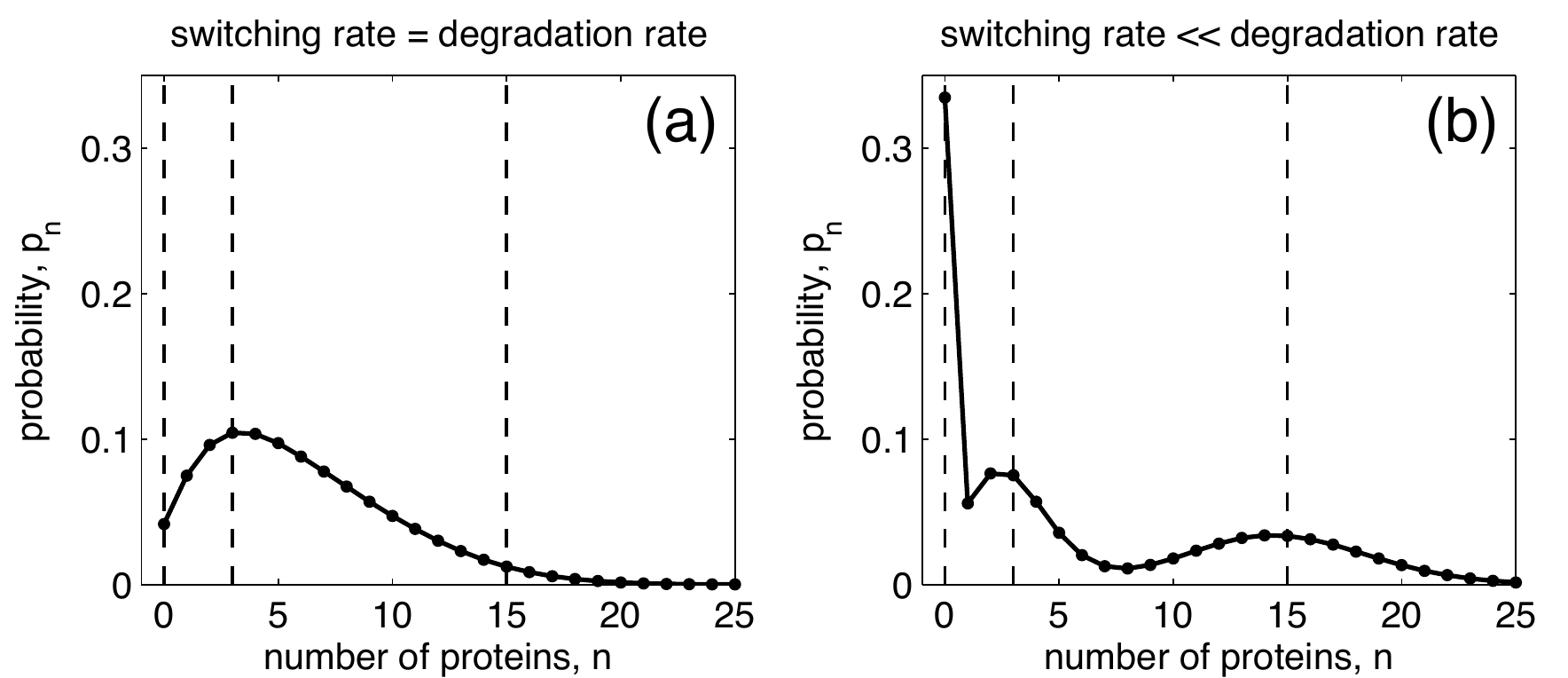}
\caption{Steady state probability distribution for transcriptional bursting with $Z=3$ states, when the transition rate among states is (a) equal to the degradation rate ($\omega = 1$), and (b) much less than the degradation rate ($\omega = 0.01$).  Production rates are $g_z = (0,3,15)$ and the rate of transition $\omega$ between any two states is equal (i.e. all off-diagonal elements of $\Omega_{zz'}$ are $\omega$ and all three diagonal elements are $-2\omega$).}
\label{fig:3state}
\end{figure}

\subsubsection{Bursting and auto-regulation}\label{burstauto}

Here we consider the simplest model that combines bursting with auto-regulation \cite{Hornos}.  The solution follows that of Sec.\ \ref{on/off}.
The gene has two states, an active state in which the regulatory protein is bound ($p_n^+$) and an inactive state in which the regulatory protein is unbound ($p_n^-$).  The regulation is incorporated by making the rate at which the gene switches to the active state proportional to the protein number:
\beq
\label{Omega2n}
\Omega_{zz'} = \begin{pmatrix} -\omega_- & \omega_+n \\ \omega_- & -\omega_+n \end{pmatrix}.
\eeq
In this case the analog of Eqn.\ \ref{G2state} for the generating function is
\beq
\label{Gautoburst}
0 = -(x-1)(\partial_x - g_\pm) G_\pm \pm \omega_+ x\partial_x G_- \mp \omega_- G_+,
\eeq
As in Sec.\ \ref{on/off}, the $+$ and $-$ equations can be combined to give a second-order differential equation, e.g. for $G_-(x)$,
\beq
0 = \partial^2_x G_- + C_1(x) \partial_x G_- + C_0(x) G_-,
\eeq
with
\beqn
C_1(x) &\equiv& \frac{g_- + g_+ + \omega_- + \omega_+ + 1 - x[g_+(1+\omega_+) + g_-]}
	{(1+\omega_+)x - 1}, \\
C_2(x) &\equiv& \frac{g_- g_+ x - g_-(g_+ + \omega_- + 1)}{(1+\omega_+)x - 1},
\eeqn
whose solution $G_-(x) = N_- e^{g_+ (x-1)} \Phi[\alpha,\beta;u]$ is proportional to the confluent hypergeometric function of the first kind (Eqn.\ \ref{hyper}) with
\beqn
\alpha &=& 1+\frac{\omega_-}{1+\omega_+}\left( 1 + \frac{\omega_+g_-}{g_- - (1+\omega_+)g_+} \right), \\
\beta &=& 1 + \frac{\omega_-}{1+\omega_+} + \frac{\omega_+g_-}{(1+\omega_+)^2}, \\
u &=& - \frac{[g_+(1+\omega_+) - g_-][(1+\omega_+)x-1]}{(1+\omega_+)^2}.
\eeqn
Again $N_-$ is set by normalization; $G_+(x)$ is computed from $G_-(x)$ using Eqn.\ \ref{Gautoburst} (bottom signs).

In Appendix \ref{Hillderiv} we consider an extension of this model, in which the transition rate to the active state is proportional to $n^h$, where $h$ is the cooperativity of the binding.  We show that in the limit of fast transitions this model yields an effective auto-regulation function of the Hill form (Eqn.\ \ref{regfun}).  Full solution of this model using the method presented here for $h=0$ and $h=1$ is of limited utility for $h>1$, because the high-order differential equation is not solvable.

\subsection{Translational bursts}

We now turn our attention to translational bursts, in which a single mRNA transcript produces many proteins in a short amount of time.  In this case the model includes just one protein production rate, but each production event results in an instantaneous burst of multiple proteins instead of one.
This approach does not explicitly consider mRNAs; instead it models the translational step as an effective burst of proteins.

With production occurring in bursts of size $N$, the birth--death master equation becomes
\beq
\label{mastbusrts}
\dot{p}_n = -g p_n - np_n + gp_{n-N} + (n+1)p_{n+1}.
\eeq
This equation is easily solved by spectral decomposition.  In terms of raising and lowering operators (Sec.\ \ref{sec:oprep}), Eqn.\ \ref{mastbusrts} reads
\beqn
\dot{\ket{G}} &=& \left[ -g - \ad\al + g{\adN} + \al \right] \ket{G} \\
\label{bursts_op}
&=& \left(-\bp\bl +\Gamma \right) \ket{G},
\eeqn
where $\bd = \ad - 1$ and $\bl = \al - g$ as before, and $\Gamma \equiv {\adN}- \ad$.
Just as in Sec.\ \ref{sec:oprep}, Eqn.\ \ref{bursts_op} benefits from a spectral expansion in the eigenfunctions $\ket{j}$ of the simple birth--death process, i.e.
\beq
\ket{G} = \sum_jG_j\ket{j},
\eeq
where
\beq
\label{eigen2}
\bp\bl \ket{j} = j\ket{j}.
\eeq
Eqn.\ \ref{bursts_op} then becomes an equation for the expansion coefficients:
\beq
\label{bbEoM}
\dot{G}_j = -jG_j + \sum_{j'} \bra{j}\Gamma\ket{j'}G_{j'}.
\eeq
The simplification
\beqn
 \bra{j}\Gamma\ket{j'}
	&=& \bra{j} \left[ \left(\bp+1\right)^N - \left(\bp+1\right) \right] \ket{j'} \\
	&=& \bra{j} \left[ \sum_{\l=0}^N \choo{N}{\l} \left(\bp\right)^\l - \left(\bp+1\right) \right] \ket{j'} \\
	&=& \bra{j} \left[ (N-1)\bd + \sum_{\l=2}^N \choo{N}{\l} \left(\bp\right)^\l \right] \ket{j'} \\
	&=& (N-1)\delta_{j,j'-1} + \sum_{\l=2}^N \choo{N}{\l} \delta_{j,j'+\l}.
\eeqn
gives
\beq
\label{bbEoM2}
\dot{G}_j = -jG_j + (N-1)G_{j-1} + \sum_{\l=2}^N \choo{N}{\l} G_{j-\l} ,
\eeq
The equation is lower-triangular, which makes a recursive solution simple.
Note that setting $N=1$ recovers Eqn.\ \ref{Gjeig} for the simple birth--death process.
The probability distribution is retrieved by inverse transform,
\beq
p_n = \sum_j G_j \ip{n}{j}.
\eeq
with $\ip{n}{j}$ computed as in Appendix \ref{app:overlap}.

Translational bursts increase noise relative to the simple birth--death process.  For example, we can easily compute the variance and mean of the steady state distribution for translational bursts (see Appendix \ref{Appendix_meansofBD}); their ratio is
\beq
\frac{\sigma^2}{\bar{n}} = \frac{1+N}{2},
\eeq
which for $N>1$ is (potentially much) greater than the birth--death result $\sigma^2/\bar{n} = 1$.

We note that within the Langevin approach we can model translational bursts as:
\beqn
\frac{d n}{d t}&=& Ng-n +\eta(t)\\
\avg{\eta(t) \eta(t')}&=&\delta(t-t') \left( Ng+n\right).
\eeqn
We obtain a power spectrum of the form:
\beq
{\cal{N}}=\frac{N+1}{\omega^2+1},
\eeq
which again is a signature of a more noisy process.

\section{Two genes}\label{twogenes}

At this point we have discussed models of a single gene, including the simple birth--death process, auto-regulation, and bursts.  Here we extend the discussion to include multiple genes with regulation.  We focus on the case of two genes, and we describe how the method may be extended to larger regulation networks.  We refer the reader to the literature for more detailed discussion of the application of the spectral method to cascades \cite{mtv} and to systems with both regulation and bursts \cite{cspan}.

Here we consider two genes, the first with protein number $n$, and the second with protein number $m$.  The first gene regulates its own expression via arbitrary function $g_n$, and it regulates the expression of the second gene via arbitrary function $q_n$.  This system can be thought of as the simplest possible regulatory network; it is also the minimum system necessary for completely describing a regulatory cascade of arbitrary length \cite{mtv}.  Because the model is general, the two degrees of freedom could correspond to two species of proteins, or to a protein and an mRNA species.

The master equation for this system is
\beqn
\label{mastern}
\dot{p}_{nm} &=& -g_np_{nm} - np_{nm} + g_{n-1}p_{n-1,m}+(n+1)p_{n+1,m}\nonumber\\
&&	+\rho \left[ -q_np_{nm} - mp_{nm} + q_np_{n,m-1}+(m+1)p_{n,m+1} \right],
\eeqn
where time is rescaled by the degradation rate of the first gene and $\rho$ is the ratio of the degradation rate of the first gene to that of the second.

\subsection{Deterministic model}

As always, the kinetic rate equations can be obtained by deriving the mean dynamics from the master equation.  Summing the master equation over both indices against either $n$ or $m$ (and performing index shifts as in Appendix \ref{app:summing}) gives
\beqn
\frac{d\bar{n}}{dt} &=& g(\bar{n}) - \bar{n}, \\
\frac{d\bar{m}}{dt} &=& q(\bar{n}) - \bar{m},
\eeqn
respectively.  Here we have assumed that $\sum_n g_n p_n \approx g(\bar{n})$ and $\sum_n q_n p_n \approx q(\bar{n})$ by taking $n = \bar{n} + \eta$ and keeping the first terms of the Taylor expansions $g(n) = g(\bar{n}) + \eta g'(\bar{n}) + \dots$ and $q(n) = q(\bar{n}) + \eta q'(\bar{n}) + \dots$, respectively.

Solving for the steady state of the deterministic system requires finding the roots of two coupled nonlinear equations, which can only be done numerically.
Solving the deterministic system is essential for finding the power spectrum and calculating the correlation functions in the Langevin approach. However, as we show in the next section, the full stochastic description can be solved quite efficiently using the spectral method, i.e. by expanding in the eigenfunctions of the simple birth--death process.

\subsection{The spectral method}
Although it is usually not possible to find an exact solution at either the master and Fokker--Planck equation levels, here we show that efficient numerical computation of the full steady state distribution is possible by expanding in the eigenfunctions of the simple birth--death process (Sec.\ \ref{sec:oprep}).
This approach is general and does not depend on the algebraic forms of the regulation functions. 

As in Sec.\ \ref{sec:oprep} we define the generating function $\ket{G} = \sum_{nm}p_{nm}\ket{n,m}$, which makes the master equation
\begin{eqnarray}
\label{eom2genes}
\dot{\ket{G}}&=&-\Lop\ket{G},
\end{eqnarray}
where the full linear operator
\beq
\Lop = \bp_n \bnn+\rho \bp_m \bmn
\eeq
is written in terms of raising and lowering operators defined analogously to Eqns.\ \ref{eq:b+}-\ref{eq:b-}:
\beqn
\label{b1}
\bp_n &=& \ad_n-1,\\
\label{b2}
\bp_m &=& \ad_m-1,\\
\label{b3}
\bnn &=& \al_n - \gh,\\
\label{b4}
\bmn &=& \al_m - \qh.
\eeqn
The difference here is that both lowering operators are $n$-dependent because of the regulation functions.
This suggests that we separate the full operator as
$\Lop = \Lop_0+\Lop_1(n)$,
where we have defined a constant part,
\beq
\Lop_0 \equiv
\bp_n\bbar_n+\rho\bp_m\bbar_m,
\eeq
with $\bbar_n \equiv \al_n - \gb$ and $\bbar_m \equiv \al_m - \qb$,
and an $n$-dependent part
 \beq
\Lop_1(n) \equiv \bp_n\Gah + \rho\bp_m\Deh,
 \eeq
with $\Gah \equiv \gb-\gh$ and $\Deh \equiv \qb-\qh$.  The operator $\Lop_0$ describes two independent birth-death processes with constant production rates $\gb$ and $\qb$ respectively, and the operator $\Lop_1$ captures how the regulated processes differ from the constant rate processes.  The constants $\gb$ and $\qb$ act as gauges: they can be chosen freely, and the final solution is independent of this choice.

We have already solved for the eigenfunctions $\ket{j,k}$ of the constant rate birth--death processes in Sec.\ \ref{sec:oprep}. The spectral method exploits this fact by expanding the solution of the full problem in these eigenfunctions:
 \beq
\label{G_1}
\ket{G} = \sum_{jk} G_{jk} \ket{j,k}.
\eeq
Eqn.\ \ref{eom2genes} then gives an algebraic relation among the expansion coefficients:
\beq
\label{jk_temp}
\dot{G}_{jk} = -(j+\rho k)G_{jk}
- \sum_{j'} \Gamma_{j-1,j'} G_{j'k}
- \rho\sum_{j'} \Delta_{jj'} G_{j',k-1},
\eeq
where
\beqn
\label{Gammajj}
\Gamma_{jj'}&=&\bra{j}\Gah\ket{j'}=\sum_{n}\ip{j}{n}(\gb-g_n)\ip{n}{j'},\\
\Delta_{jj'}&=&\bra{j}\Deh\ket{j'}=\sum_{n}\ip{j}{n}(\qb-q_n)\ip{n}{j'}.
\eeqn
Eqn.\ \ref{jk_temp} enjoys the property that it is sub-diagonal in the second gene's eigenvalue $k$.  This feature comes from the fact that both regulation functions are dependent on the first gene's protein number $n$ (and not $m$).  The sub-diagonal structure allows the matrix $G_{jk}$ to be computed iteratively by columns in $k$, which makes the method numerically efficient.
The initial condition
\beq
\label{Ginit}
G_{j0}=\ip{j,k=0}{G}=\sum_{nm}p_{nm}\ip{j}{n}\ip{0}{m} = \sum_np_n\ip{j}{n}
\eeq
requires the marginal probability distribution of the first gene $p_n$, which can be found recursively (Eqn.\ \ref{pn}) as described in Sec.\ \ref{automaster} on one-dimensional auto regulation.  The joint probability distribution is retrieved by inverse transform:
\beq
p_{nm} = \sum_{jk} G_{jk} \ip{n}{j}\ip{m}{k}.
\eeq

\begin{figure}
\includegraphics[scale=1]{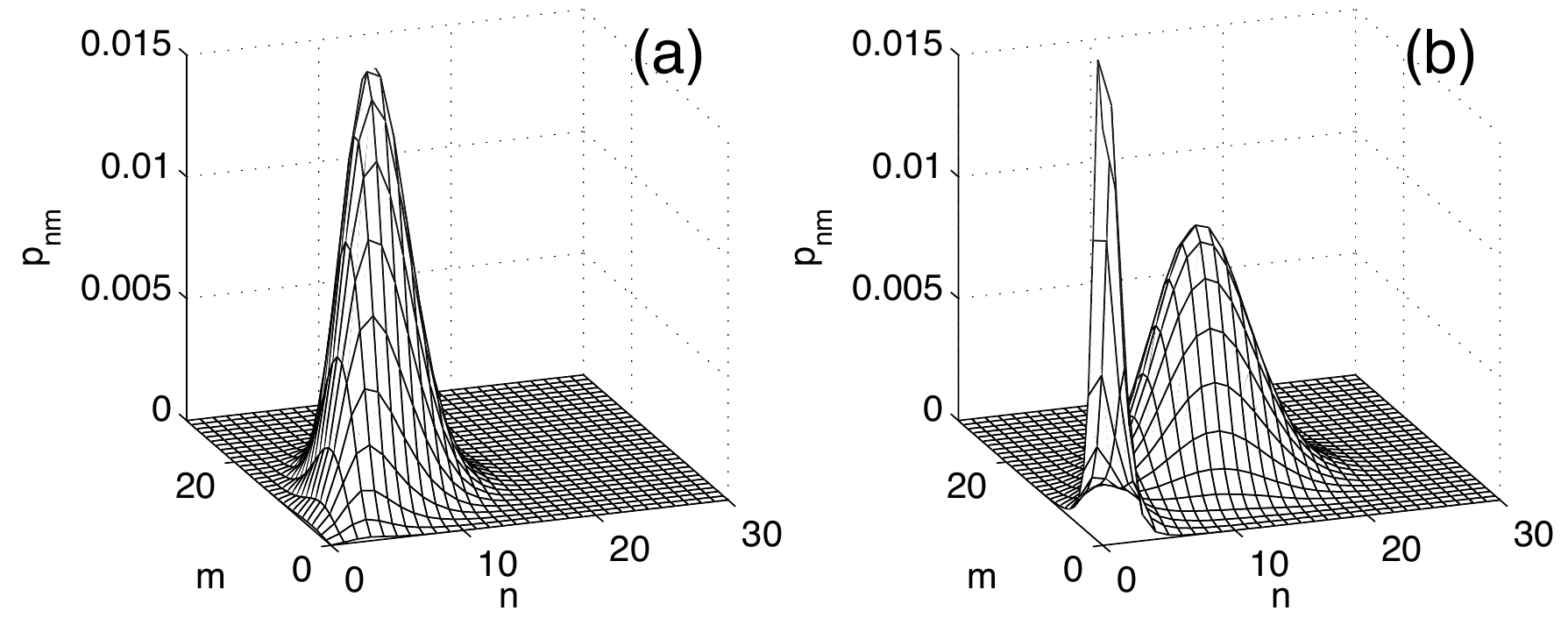}
\caption{Demonstration of the spectral method with two genes.  Regulation is of Hill form, $q_n = (q_-K^h+q_+n^h)/(K^h+n^h)$, with $q_- = 1$, $q_+ = 12$, $K = 4$, and $h = 4$.  In (a) the first gene undergoes a simple birth--death process with $g_n = 7$; in (b) the first gene regulates itself with $g_n = q_n$.  Degradation rates are equal: $\rho = 1$.}
\label{joints}
\end{figure}

\begin{figure}
\includegraphics[scale=1]{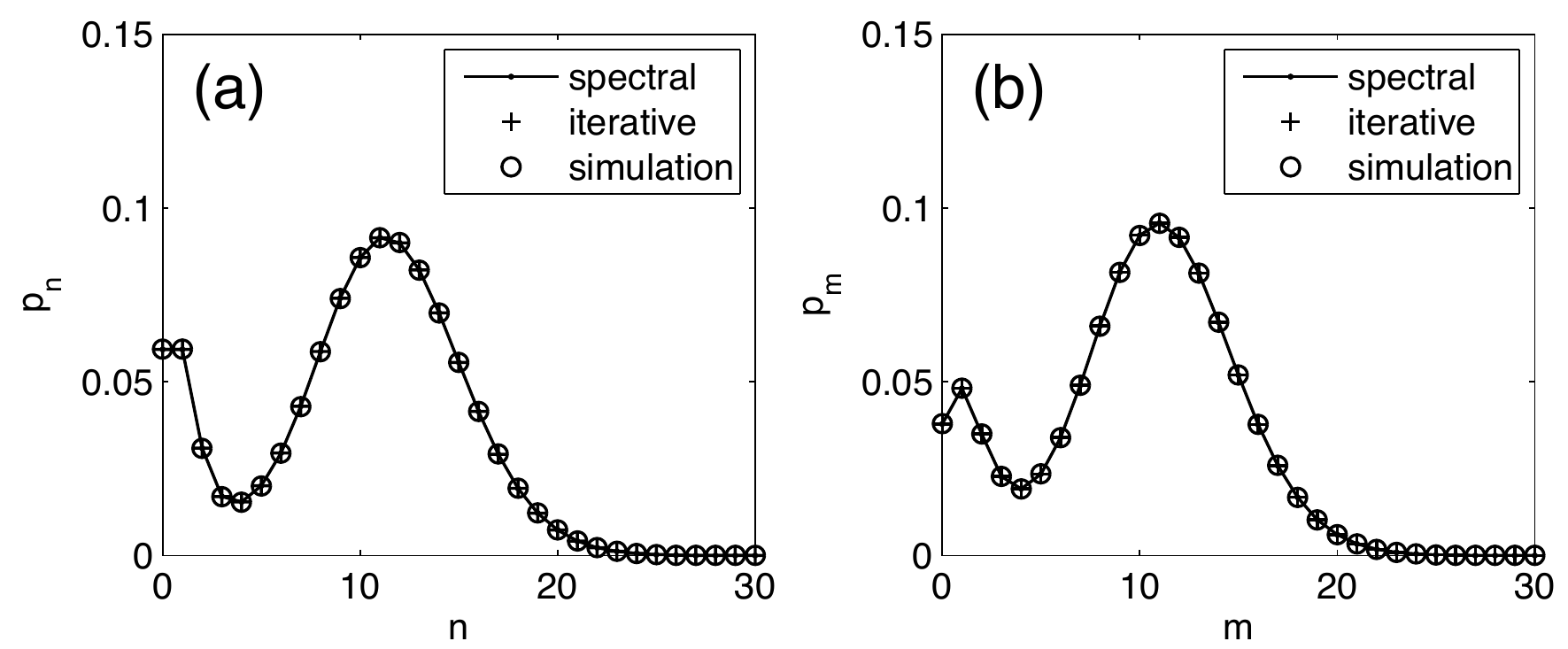}
\caption{Marginal distributions for the (a) first and (b) second gene in a two-gene network, summed from the joint distribution in Fig \ref{joints}(b).  Agreement is demonstrated among distributions obtained by the spectral method (solid), by direct iterative solution of the master equation (Eqn.\ \ref{mastern}; plus signs),  and by standard Gillespie simulation \cite{Gillespie} (circles).}
\label{marginals}
\end{figure}

To demonstrate the capability of the spectral method, Fig.\ \ref{joints} plots the steady state joint probability distribution for a particular choice of regulation function, for cases (a) without and (b) with auto-regulation.  To demonstrate the accuracy of the spectral method, Fig.\ \ref{marginals} compares steady state marginal distributions obtained by the spectral method with those obtained both by direct iterative solution of Eqn.\ \ref{mastern} and by standard Gillespie simulation \cite{Gillespie}.  The spectral method attains excellent agreement with both methods in orders of magnitude less runtime than either other method; full details on the efficiency and accuracy of the spectral method are available in \cite{mtv}.

The spectral decomposition presented here constitutes an expansion in only one of many possible eigenbases: that with constant production rates. Other spectral expansions are possible, as discussed in detail in \cite{cspan}; we do not repeat the discussion here.

\section{Summary}

In this pedagogical chapter, we have attempted to introduce the reader to the phenomenon of intrinsic noise and to the simplest possible mathematical tools for modeling. We have avoided detailed discussion of a vast literature in Monte Carlo simulation of such phenomena in favor of simple calculations which can be performed analytically or which are useful in numerical solutions of the master equation itself (rather than random generation of sample trajectories).

The discussion also hopefully illustrates how different problem settings motivate different analytic methods, many of which have close parallels in the physics canon. These include approximation of differences with derivatives, yielding a description very similar to that used in Fokker-Planck descriptions; as well as the algebraic relations among the different eigenfunctions of the birth-death process itself, yielding an algebra of ``ladder operators" similar to that used in quantum mechanics. 

While the literature in simulation of intrinsic noise has grown immensely since the introduction of time-varying Monte Carlo methods over thirty years ago \cite{Gillespie}, it is our hope that for model systems these calculations  will help point out how the linearity of master equations may be exploited to yield either solutions, numerical methods, or analytical approximations which give more direct insight into the deign principles of stochastic regulatory networks.

\appendix

\section{Deriving the master equation}\label{app:master}
Here we derive the master equation in one dimension from the laws of probability.
The probability of having $n$ proteins at a time $t+\tau$ (where $\tau$ is small) is equal to the probability of having had $n'$ proteins at time $t$ multiplied by the (possibly time-dependent) probability of transitioning from $n'$ to $n$ in time $\tau$, summed over all $n'$:
\beq
p_n(t+\tau) = \sum_{n'} p_{n'}(t) p_{n|n'}(t,\tau).
\label{cond_meder}
\eeq
The transition probability $p_{n|n'}(t,\tau)$ has two contributions: (i) the probability of transitioning from state $n'\neq n$ to state $n$, equal to a transition rate $w_{nn'}(t)$ times the transition time $\tau$, and (ii) the probability of starting and remaining in state $n'=n$, which we call $\pi_n(t)$:
\beq
p_{n|n'}(t,\tau) = (1-\delta_{nn'}) w_{nn'}(t) \tau + \delta_{nn'} \pi_n(t) 
\label{trans_meder}
\eeq
Applying the normalization condition  $\sum_n p_{n|n'}(t,\tau)=1$ to Eqn.\ \ref{trans_meder}, we obtain
\beq
\pi_n(t) = 1-\tau \sum_{n'\neq n} w_{n'n}(t),
\label{a_meder}
\eeq
making Eqn.\ \ref{cond_meder}
\beqn
p_n(t+\tau) &=& \sum_{n'} p_{n'}(t) \left[ (1-\delta_{nn'}) w_{nn'}(t) \tau +
	\delta_{nn'} \left( 1-\tau \sum_{n''\neq n} w_{n''n}(t) \right) \right] \\
&=& \tau \sum_{n'} p_{n'}(t) w_{nn'}(t) - \tau p_n(t)w_{nn}
	+ p_n(t) - \tau p_n(t) \sum_{n''\neq n} w_{n''n}(t) \\
&=& p_n(t) + \tau \left[ \sum_{n'} p_{n'}(t) w_{nn'}(t) - p_n(t) \sum_{n'} w_{n'n}(t) \right].
\label{cond2_meder}
\eeqn
Taking the limit
$\tau \rightarrow 0$, we obtain
 \beq
\frac{d p_n}{d t} = \sum_{n'} p_{n'} w_{nn'}(t) - p_n \sum_{n'} w_{n'n}(t),
\eeq
as in Eqn.\ \ref{eq:mastergeneral}.

\section{Summation of the birth--death master equation}\label{app:summing}

Here we sum the one-dimensional birth--death master equation over $n$ against (i) $n$, to derive the mean dynamics, (ii) $x^n$, to derive the equation of motion in generating function space, and (iii) $\ket{n}$, to find the equation of motion in operator notation.

First we sum the master equation (Eqn.\ \ref{eq:master}) over $n$ against $n$ to derive the dynamics of the mean $\avg{n} = \sum_n np_n$:
\beq
\partial_t \sum_n n p_n = -g \sum_n n p_n - \sum_n n^2 p_n
	+ g \sum_n n p_{n-1} + \sum_n n (n+1) p_{n+1}.
\eeq
The left-hand side (LHS) is the time derivative of the mean.  On the right-hand side (RHS), recalling that $n$ runs from $0$ to $\infty$, we may define for the third term $n'\equiv n-1$ and for the fourth term $n''\equiv n+1$, giving
\beq
\label{shift1}
\partial_t \avg{n} = -g \sum_n n p_n - \sum_n n^2 p_n
	+ g \sum_{n'=-1}^\infty (n'+1) p_{n'} + \sum_{n''=1}^\infty (n''-1) n'' p_{n''}.
\eeq
The lower limits on all sums can be set to $0$ once more without changing the expression (in the third term we impose $p_{-1} \equiv 0$ since protein number cannot be negative); simplifying gives
\beqn
\partial_t \avg{n} &=& -g \sum_n n p_n - \sum_n n^2 p_n
	+ g \sum_{n'} n' p_{n'} + g \sum_{n'} p_{n'}
	+ \sum_{n''} n''^2 p_{n''} - \sum_{n''} n'' p_{n''} \\
&=& g \sum_{n'} p_{n'} - \sum_{n''} n'' p_{n''} \\
&=& g - \avg{n},
\eeqn
as in Eqn.\ \ref{avgn}.

Next we sum the master equation (Eqn.\ \ref{eq:master}) over $n$ against $x^n$ to derive the equation of motion of the generating function $G(x,t) = \sum_n p_n(t) x^n$:
\beq
\partial_t \sum_n p_n x^n = -g \sum_n p_n x^n - \sum_n n p_n x^n
	+ g \sum_n p_{n-1} x^n + \sum_n (n+1) p_{n+1} x^n.
\eeq
The LHS and the first term on the RHS reduce directly to functions of $G$.  The third and fourth terms on the RHS benefit from the same index shifts applied to obtain Eqn.\ \ref{shift1}, giving
\beq
\partial_t G = -g G - \sum_n p_n n x^n
	+ g \sum_{n'} p_{n'} x^{n'+1} + \sum_{n''} n'' p_{n''} x^{n''-1}.
\eeq
We may now eliminate bare appearances of $n$ in the second and fourth terms on the RHS using $n x^n = x\partial_x x^n$ and $n x^{n-1} = \partial_x x^n$ respectively, giving
\beqn
\partial_t G &=& -g G - \sum_n p_n x\partial_x x^n
	+ g \sum_{n'} p_{n'} x^{n'+1} + \sum_{n''} p_{n''} \partial_x x^{n''} \\
&=& -gG - x\partial_xG + gxG + \partial_xG \\
&=& -(x-1)(\partial_x - g)G
\eeqn
as in Eqn.\ \ref{eq:Gss}.

Finally we sum the master equation (Eqn.\ \ref{eq:master}) over $n$ against $\ket{n}$ to derive the equation of motion of the generating function in operator notation $\ket{G} = \sum_n p_n \ket{n}$:
\beq
\label{mastersumket}
\partial_t \sum_n p_n \ket{n} = -g \sum_n p_n \ket{n} - \sum_n n p_n \ket{n}
	+ g \sum_n p_{n-1} \ket{n} + \sum_n (n+1) p_{n+1} \ket{n}.
\eeq
Again we may reduce the LHS and the first term on the RHS directly, and we may apply index shifts to the third and fourth terms on the RHS:
\beq
\partial_t \ket{G} = -g \ket{G} - \sum_n np_n \ket{n}
	+ g \sum_{n'} p_{n'} \ket{n'+1} + \sum_{n''} n'' p_{n''} \ket{n''-1}.
\eeq
Now defining operators $\ad$ and $\al$ that raise and lower the protein number by 1 respectively, i.e.\
\begin{eqnarray}
\label{eq:a+right2}
\hat{a}^+|n\rangle &\equiv& |n+1\rangle, \\
\label{eq:a-right2}
\hat{a}^-|n\rangle &\equiv& n|n-1\rangle,
\end{eqnarray}
Eqn.\ \ref{mastersumket} can be written
\beqn
\partial_t \ket{G} &=& -g \ket{G} - \sum_n p_n \ad\al \ket{n}
	+ g \sum_{n'} p_{n'} \ad \ket{n'} + \sum_{n''} p_{n''} \al \ket{n''} \\
&=& -g\ket{G} - \ad\al\ket{G} + g\ad\ket{G} + \al\ket{G} \\
&=& -(\ad-1)(\al-g)\ket{G}
\eeqn
as in Eqn.\ \ref{eom_op}.

\section{
Statistics of the birth--death distribution}\label{Appendix_meansofBD}

In this appendix we calculate the means of a birth--death distribution. In the simplest case, starting directly from the solution in Eqn.\ \ref{eq:pstar1} we obtain:
\beqn
\avg{n}&=&\sum_n n p_n= \sum_n n \frac{g^n}{n!} e^{-g}=e^{-g} g \partial_g \sum_n \frac{g^n}{n!} =g.
\eeqn
Similarly we can calculate
\beqn
\avg{n(n-1)}&=& \sum_n n(n-1) \frac{g^n}{n!} e^{-g}=e^{-g} g^2 \partial^2_g \sum_n \frac{g^n}{n!} =g^2,
\eeqn
which gives us the variance
\beq
\avg{n^2}-\avg{n}^2=g^2+g-g^2=g.
\eeq

These results can also be obtained in the operator formalism, and the fact that for steady state $\ket{G}=\ket{j=0}$. For example:
\beq
\avg{n}=\bra{x}\ad \al \ket{j=0}|_{x=1}=\bra{x}\ad \al \ket{G}|_{x=1}=\bra{x}\ad \al \sum_n p_n \ket{n}|_{x=1}=\sum_n n p_n x^n |_{x=1}=\sum_n n p_n.
\eeq
To use the number operator on the $\ket{j=0}$ state we need to rewrite it in terms of $\ad=\bp+1$ and $\al=\bl+g$. Acting to the left we obtain:
\beq
\avg{n}=\bra{x} \left(\bp\bl+g+\bl+g\bp \right)\ket{0}_{x=1}= \bra{x} g \ket{0}_{x=1}+ \bra{x} g \ket{1}_{x=1}=g.
\eeq

We can also calculate the mean of the bursty process. Multiplying Eqn.\ \ref{mastbusrts} by $n$ and summing over $n$, we use,
\beq
\sum_{n=0}^{\infty} n p_{n-N}=\sum_{n=0}^{\infty} (n+N)p_n=\avg{n}+N,
\eeq
to obtain:
\beq
g(\avg{n}+N)-g\avg{n}+\avg{n^2}-\avg{n}-\avg{n^2}=0,
\eeq
which is solved by:
\beq
\avg{n}=gN.
\eeq
The second moment $\avg{n^2}=\sum n^2 p_n$ is calculated the same way:
\beq
\avg{n^2}=\frac{1}{2}\left( 2g\avg{n} N+\avg{n}+gN^2\right)=(gN)^2+\frac{gN^2+gN}{2},
\eeq
which yields the variance as
\beq
\delta n^2=\avg{n^2}-\avg{n}^2=\frac{gN}{2} (1+N).
\eeq

\section{Asymptotic distributions for large protein number}\label{app:gauss}

Here we show that in the limit of large protein number $n$, the steady state of the birth--death process (the Poisson distribution, Eqn.\ \ref{eq:pstar1}) approaches a Gaussian distribution with mean and variance equal to $g$, the ratio of production to degradation rates.  We then show that in the same limit, the steady state solution to the birth--death Fokker--Planck equation (Eqn. \ref{FPssbd}) also approaches a Gaussian distribution with mean and variance equal to $g$.

The large-$n$ limit of the Poisson distribution is most conveniently evaluated by first taking the log, which allows one to make use of Stirling's approximation, $\log n! \approx n\log n - n + \log \sqrt{2\pi n}$ for large $n$:
\beq
\log p(n) \approx -g + n\log g - n\log n + n - \log \sqrt{2\pi n}.
\eeq
The derivative
\beq
\partial_n \log p(n) = \log \frac{g}{n} + {\cal O}(1/n)
\eeq
vanishes at the maximum
\beq
n = g
\eeq
about which we Taylor expand to second order:
\beqn
\log p(n) &\approx& \log p(g) + \frac{1}{2}(n-g)^2\partial_n^2 \left[ \log p(n) \right]_g \\
&=& -\log \sqrt{2\pi g} - \frac{(n-g)^2}{2g}.
\eeqn
The result is a Gaussian distribution with mean and variance equal to $g$:
\beq
p(n) = \frac{1}{\sqrt{2\pi g}} e^{-(n-g)^2/(2g)}.
\eeq

The large-$n$ limit of the Fokker--Planck distribution is similarly evaluated: the log and it's derivatives are
\beqn
\log p(n) &=& \log \frac{N}{g} + (4g-1)\log \left( 1+\frac{n}{g} \right), \\
\partial_n \log p(n) &=& \frac{4g-1}{n+g} - 2, \\
\partial_n^2 \log p(n) &=& \frac{1-4g}{(n+g)^2}.
\eeqn
The first derivative vanishes at $n = g-1/2 \approx g$ for large $g$, at which the second derivative evaluates to $\partial_n^2 \left[ \log p(n) \right]_g = -1/g + 1/(4g^2) \approx -1/g$.  Taylor expanding to second order and exponentiating then give
\beq
p(n) = \frac{1}{\sqrt{2\pi g}} e^{-(n-g)^2/(2g)},
\eeq
where $N$ is eliminated by normalization.

\section{Orthonormality and the inner product}\label{app:nbasis}
Here we show that interpreting the generating function as a Fourier transform motivates a particular choice of inner product and conjugate state in the protein number basis.

We have defined the generating function in terms of a continuous variable $x$ as $G(x) = \sum_n p_n x^n$.  We could equivalently write $x\equiv e^{i\kappa}$ and define
\begin{equation}
\label{eq:Gkdef}
G(\kappa) = \sum_n p_n e^{in\kappa},
\end{equation}
which makes clear that the generating function is simply the Fourier transform of the probability distribution in protein number $n$.
In the state notation commonly used in quantum mechanics \cite{sakurai1985modern}, Eqn.\ \ref{eq:Gkdef} is written $\ip{\kappa}{G} = \sum_n p_n \ip{\kappa}{n}$, where $\langle \kappa|G \rangle \equiv G(\kappa)$ and
\begin{equation}
\label{kstate}
\langle \kappa|n\rangle \equiv e^{in\kappa}.
\end{equation}

Eqn.\ \ref{kstate} is the representation of $\ket{n}$ in $\kappa$ space.  In order to compute projections of $\ket{n}$ onto other states or itself using this representation, we must define a conjugate state $\ip{n}{\kappa}$ and a consistent inner product.  With complex exponentials, it is common to make the conjugate state the complex conjugate,
\begin{equation}
\ip{n}{\kappa} \equiv \ip{\kappa}{n}^* = e^{-in\kappa},
\end{equation}
and the inner product the integral
\beq
\ip{f}{h} \equiv \int_0^{2\pi} \frac{d\kappa}{2\pi} \ip{f}{\kappa}\ip{\kappa}{h},
\eeq
for any $f$ and $h$.
Given the definition of conjugate state, this choice of inner product ensures the orthonormality of the $\ket{n}$ states:
\begin{equation}
\label{eq:kip}
\ip{n}{n'} = \int_0^{2\pi} \frac{d\kappa}{2\pi} \langle n|\kappa\rangle\langle \kappa|n'\rangle
	= \int_0^{2\pi} \frac{d\kappa}{2\pi} e^{i(n'-n)\kappa}
	= \delta_{nn'}.
\end{equation}

We may now reinterpret these definitions in terms of our original variable $x$.  Since $x=e^{i\kappa}$ we have $dx = ie^{i\kappa}d\kappa = ixd\kappa$,
and in the orthonormality condition (Eqn.\ \ref{eq:kip}) the integration from $0$ to $2\pi$ along the real $k$ line becomes contour integration along the unit circle in the complex $x$ plane:
\begin{equation}
\label{eq:xiptemp}
\ip{n}{n'}
	= \int_0^{2\pi} \frac{dk}{2\pi} e^{-ink} e^{in'k}
	= \oint \frac{dx}{2\pi ix} x^{-n} x^{n'}
	= \oint \frac{dx}{2\pi i} \frac{1}{x^{n+1}} x^{n'}
	= \delta_{nn'}.
\end{equation}
Since we have defined
\beq
\ip{x}{n} = x^n,
\eeq
Eqn.\ \ref{eq:xiptemp} suggests the definition of conjugate state
\begin{equation}
\label{eq:nonxapp}
\langle n|x\rangle = \frac{1}{x^{n+1}}
\end{equation}
and inner product
\begin{equation}
\label{eq:ip}
\langle f|h\rangle = \oint \frac{dx}{2\pi i} \langle f|x\rangle\langle x|h\rangle
\end{equation}
in $x$ space.

With these definitions, we will find Cauchy's theorem,
\begin{equation}
\label{eq:cauchy}
\oint \frac{dx}{2\pi i} \frac{f(x)}{(x-a)^{n+1}} = \frac{1}{n!} \partial^n_x \left[ f(x) \right]_{x=a} \theta(n),
\end{equation}
(with the convention $\theta(0) = 1$ for the Heaviside function) very useful in evaluating projections.  For example, we may immediately use it to confirm the orthonormality condition,
\begin{equation}
\label{eq:nn}
\langle n|n'\rangle = \oint \frac{dx}{2\pi i} \frac{1}{x^{n+1}} x^{n'}
	= \frac{1}{n!} \partial^n_x \left[ x^{n'} \right]_{x=0} \theta(n)
	= \delta_{nn'},
\end{equation}
and in Appendix \ref{app:overlap} we use it to compute analytic expressions for the projections between protein number states $\ket{n}$ and birth--death eigenstates $\ket{j}$.

\section{Properties of the raising and lowering operators}\label{app:aproperties}

Just as in the operator treatment of the quantum harmonic oscillator \cite{sakurai1985modern}, in this paper we have defined operators $\ad$ and $\al$ that act on $\ket{n}$ states by raising and lowering the protein number $n$ by 1 respectively, i.e.\
\begin{eqnarray}
\label{eq:a+right3}
\hat{a}^+|n\rangle &=& |n+1\rangle, \\
\label{eq:a-right3}
\hat{a}^-|n\rangle &=& n|n-1\rangle
\end{eqnarray}
(note, however, that the prefactors here, $1$ and $n$, are different than those conventionally used for the harmonic oscillator, $\sqrt{n+1}$ and $\sqrt{n}$, respectively).  In this appendix we derive the actions of these operators when acting to the left, as well as their commutation relation.

The actions of $\ad$ and $\al$ to the left can be found by projecting onto Eqns.\ \ref{eq:a+right3} and \ref{eq:a-right3} the conjugate state $\bra{n'}$:
\begin{eqnarray}
\label{aleftapp1}
\bra{n'}\ad\ket{n} &=& \ip{n'}{n+1} = \delta_{n',n+1} = \delta_{n'-1,n} = \ip{n'-1}{n}, \\
\label{aleftapp2}
\bra{n'}\al\ket{n} &=& \bra{n'}n\ket{n-1} = n\delta_{n',n-1} = (n'+1)\delta_{n'+1,n} = (n'+1)\ip{n'+1}{n}.
\end{eqnarray}
The first and last expressions in each case imply
\begin{eqnarray}
\label{eq:a+2}
\langle n'|\hat{a}^+ &=& \langle n'-1|, \\
\label{eq:a-2}
\langle n'|\hat{a}^- &=& (n'+1)\langle n'+1|,
\end{eqnarray}
as in Eqns.\ \ref{eq:a+}-\ref{eq:a-}.

The commutation relation between $\hat{a}^-$ and $\hat{a}^+$ is defined as $\left[ \al, \ad \right] \equiv \al\ad - \ad\al$.  It is evaluated by considering its action on a state $\ket{n}$:
\beq
\left[ \al, \ad \right]\ket{n} = \al\ad\ket{n} - \ad\al\ket{n} = \al\ket{n+1} - \ad n\ket{n-1}
	= (n+1)\ket{n} - n\ket{n} = \ket{n}.
\eeq
The first and last expressions imply
\begin{equation}
\left[ \hat{a}^-, \hat{a}^+ \right] = 1,
\end{equation}
just as with the quantum harmonic oscillator.
 
\section{Eigenvalues and eigenfunctions in the operator representation}\label{app:bproperties}

Here we derive the eigenfunctions $\ket{\lambda_j}$ (also called eigenstates in the operator representation) and eigenvalues $\lambda_j$ of the operator $\Lop = \bp\bl$, for which
\beq
\label{Leigapp}
\Lop\ket{\lambda_j} = \lambda_j\ket{\lambda_j},
\eeq
as well as the actions of the individual operators $\bp$ and $\bl$ on the eigenstates.  We will find that the commutation relation
\beq
\label{bcomapp}
\left[ \bl, \bp \right] = 1
\eeq
and the existence of the steady state solution $\ip{x}{G}=e^{q(x-1)}$, for which
\beq
\label{ssapp}
\Lop\ket{G} = 0,
\eeq
are all that are necessary to completely define the eigenvalues and eigenstates of $\hat{\L}$.  The treatment here finds many parallels with the derivation of the eigenvalue spectrum of the quantum harmonic oscillator \cite{sakurai1985modern}.

First it is useful to compute the commutation relation of $\Lop$ with each of its components $\bp$ and $\bl$:
\begin{eqnarray}
\left[\hat{b}^+, \hat{\cal L}\right] &=& \left[\hat{b}^+, \hat{b}^+\hat{b}^-\right]
	= \hat{b}^+\left[\hat{b}^+, \hat{b}^-\right] + \left[\hat{b}^+, \hat{b}^+\right]\hat{b}^-
	= -\hat{b}^+, \\
\left[\hat{b}^-, \hat{\cal L}\right] &=& \left[\hat{b}^-, \hat{b}^+\hat{b}^-\right]
	= \hat{b}^+\left[\hat{b}^-, \hat{b}^-\right] + \left[\hat{b}^-, \hat{b}^+\right]\hat{b}^-
	= \hat{b}^-
\end{eqnarray}
(here we have used Eqn.\ \ref{bcomapp} and the properties $[f,gh] = g[f,h] + [f,g]h$ and $[f,f] = 0$ for any $f$, $g$, and $h$).  We now consider a particular eigenvalue $\lambda$, and evaluate the action of $\hat{\cal L} \hat{b}^+$ on $\ket{\lambda}$:
\beq
\label{Lbaction}
\Lop\bp\ket{\lambda} = \bp\Lop\ket{\lambda} - \left[ \bp, \Lop \right] \ket{\lambda}
	= \bp \lambda \ket{\lambda} + \bp \ket{\lambda}
	= (\lambda+1) \bp \ket{\lambda}.
\eeq
The equality of the first and last expressions,
\beq
\Lop \left( \bp \ket{\lambda} \right) = (\lambda+1) \left( \bp \ket{\lambda} \right)
\eeq
reveals two results: (i) the existence of a state with eigenvalue $\lambda$ implies the existence of a state with eigenvalue $\lambda+1$, and (ii) the state with eigenvalue $\lambda+1$ is proportional to $\bp\ket{\lambda}$.
By induction the first result means that the eigenvalues are spaced by $1$; moreover, the existence of a state with eigenvalue zero (the steady state solution; Eqn.\ \ref{ssapp}) anchors the eigenvalues to the integers, i.e.\
\beq
\lambda_j = j
\eeq
for integer $j$.  The second result can now be written explicitly
\beq
\label{braise}
\bp\ket{j} = \ket{j+1},
\eeq
where we are free by normalization to set the proportionality constant to $1$.  Eqn.\ \ref{braise} demonstrates that $\bp$ is a raising operator for the eigenstates $\ket{j}$.

Evaluating the action of $\Lop\bl$ on eigenstate $\ket{j}$ similarly reveals that $\bl\ket{j}$ is proportional to $\ket{j-1}$ (cf.\ Eqn.\ \ref{Lbaction}); consistency with the eigenvalue equation (Eqn.\ \ref{Leigapp}) requires that the proportionality constant be $j$, giving
\beq
\label{blower}
\bl\ket{j} = j\ket{j-1}.
\eeq
Eqn.\ \ref{blower} demonstrates that $\bl$ is a lowering operator for the eigenstates $\ket{j}$.  The operators $\bp$ and $\bl$ raise and lower $\ket{j}$ as $\ad$ and $\al$ do $\ket{n}$; therefore they act to the left as (cf.\ Eqns.\ \ref{aleftapp1}-\ref{aleftapp2})
\beqn
\bra{j}\bp &=& \bra{j-1}, \\
\bra{j}\bl &=& (j+1)\bra{j+1}.
\eeqn

Eqn.\ \ref{blower} imposes a floor on the eigenvalue spectrum, since $\bl\ket{0} = 0$, and no lower eigenstates can be generated.  Therefore the eigenvalues are limited to nonnegative integers:
\beq
j \in \{0,1,2,3,\dots\}.
\eeq

Eqn.\ \ref{braise} describes how any state can be obtained from the $j=0$ state:
\beq
\label{raisej}
\ket{j} = \left( \bp \right)^j \ket{0}.
\eeq
Recalling that $\bp = \ad-1$ and that $\ket{0}$ is the steady state solution, Eqn. \ref{raisej} can be used to derive the representation of the eigenfunctions in $x$ space, $\ip{x}{j}$.  Projecting $\bra{x}$ onto Eqn.\ \ref{raisej} and recalling that $\ad$ corresponds to $x$ (Eqn.\ \ref{correspond1}) and that $\ip{x}{0} = e^{g(x-1)}$ (Eqn.\ \ref{eq:Gform}), we obtain
\beq
\label{eigxapp}
\ip{x}{j} = (x-1)^je^{g(x-1)}.
\eeq
We may now define the conjugate state $\ip{j}{x}$ such that the $\ket{j}$ are orthonormal under the inner product defined in Eqn.\ \ref{eq:ip}:
\beq
\label{eq:jj}
\delta_{jj'} = \ip{j}{j'} = \oint\frac{dx}{2\pi i}\ip{j}{x}\ip{x}{j'}
	= \oint\frac{dx}{2\pi i}\ip{j}{x} (x-1)^{j'} e^{g(x-1)}
	= \oint\frac{dy}{2\pi i} y^{j'} f_j(y),
\eeq
where $y\equiv x-1$ and $f_j(x-1) \equiv e^{g(x-1)}\ip{j}{x}$.  Eqn.\ \ref{eq:jj} is equivalent to Eqn.\ \ref{eq:xiptemp}, from which we identify $f_j(y) = 1/y^{j+1}$ and thus
\beq
\label{eigxappcon}
\ip{j}{x} = \frac{e^{-g(x-1)}}{(x-1)^{j+1}}.
\eeq

We have shown that the operator $\Lop = \bp\bl$ has nonnegative integer eigenvalues $j$ and eigenfunctions (in $x$ space) given by Eqns.\ \ref{eigxapp} and \ref{eigxappcon}.

\section{Overlaps between protein number states and eigenstates}\label{app:overlap}

Here we describe two methods for computing the overlaps $\ip{n}{j}$ and $\ip{j}{n}$ between the protein number states $\ket{n}$ and the eigenstates $\ket{j}$: by contour integration and by recursive updating.

The first method evaluates the overlaps using the inner product defined in Eqn. \ref{eq:ip} and Cauchy's theorem (Eqn.\ \ref{eq:cauchy}).   Recalling the representations in $x$ space of the protein number states and eigenstates (Eqns. \ref{eq:xonn}, \ref{eq:nonx}, and \ref{efunc2}-\ref{efunccon}), the first overlap becomes
\beq
\ip{n}{j} = \oint \frac{dx}{2\pi i} \ip{n}{x} \ip{x}{j}
	= \oint \frac{dx}{2\pi i} \frac{e^{g(x-1)}(x-1)^j}{x^{n+1}}
	=  \frac{1}{n!} \partial^n_x \left[ e^{g(x-1)}(x-1)^j \right]_{x=0}.
\eeq
Repeated derivatives of a product follow a binomial expansion, i.e.\
\beqn
\ip{n}{j} &=& \frac{1}{n!}\sum_{\ell=0}^n \frac{n!}{\ell! (n-\ell)!}
	\partial^{n-\ell}_x\left[ e^{g(x-1)} \right]_{x=0}\partial^\ell_x\left[ (x-1)^j \right]_{x=0} \\
&=& \sum_{\ell=0}^n \frac{1}{\ell! (n-\ell)!}\
	\left[ g^{n-\ell}e^{-g} \right]\left[ \frac{j!}{(j-\ell)!} (-1)^{j-\ell} \theta(j-\ell) \right] \\
\label{nj}
&=& (-1)^j e^{-g} g^n j! \xi_{nj},
\eeqn
where we define
\beq
\label{xi}
\xi_{nj} \equiv \sum_{\ell=0}^{\min(n,j)} \frac{1}{\ell!(n-\ell)!(j-\ell)!(-g)^\ell}.
\eeq
Following similar steps, the conjugate overlap evaluates to
\beq
\label{jn}
\ip{j}{n} = n!(-g)^j\xi_{nj},
\eeq
with $\xi_{nj}$ as in Eqn.\ \ref{xi}.  For the special case $j=0$, Eqns.\ \ref{nj} and \ref{jn} reduce to Eqns.\ \ref{special1}-\ref{special2}.

It is more computationally efficient to compute the overlaps recursively using rules that can be derived from the raising and lowering operations (Eqns. \ref{eq:a+right}-\ref{eq:a-right}, \ref{eq:a+}-\ref{eq:a-}, and \ref{baction1}-\ref{baction4}). For example, using the raising operators,
\beq
\label{up1}
\ip{n}{j+1} = \bra{n}\bd\ket{j}=\bra{n}(\ad-1)\ket{j} = \ip{n-1}{j}-\ip{n}{j},
\eeq
which can be initialized using $\ip{n}{0}=e^{-g}g^n/n!$ (Eqn.\ \ref{nj}) and updated recursively in $j$. Eqn.\ \ref{up1} makes clear that in $n$ space the $(j+1)$th mode is simply the (negative of the) discrete derivative of the $j$th mode.  The lowering operators give an alternative update rule,
\beq
\label{up2}
(n+1)\ip{n+1}{j}=\bra{n}\al\ket{j}=\bra{n}(\bl+g)\ket{j} = j\ip{n}{j-1}+g\ip{n}{j},
\eeq
which can be initialized using $\ip{0}{j}=(-1)^je^{-g}$ (Eqn.\ \ref{nj}) and updated recursively in $n$.
One may similarly derive recursion relations for $\ip{j}{n}$, i.e.\
\beqn
\label{up3}
\ip{j}{n+1}&=&\ip{j-1}{n}+\ip{j}{n},\\
\label{up4}
(j+1)\ip{j+1}{n}&=&n\ip{j}{n-1}-g\ip{j}{n},
\eeqn
initialized with $\ip{j}{0}=(-g)^j/j!$ or $\ip{0}{n}=1$ respectively (Eqn.\ \ref{jn}) and updated recursively in $n$ or $j$ respectively.
Two-term recursion relations can be similarly derived from the full operator $\bp\bl$ \cite{cspan}.

\section{The equivalence of the Fokker--Planck and Langevin descriptions}\label{Appendix_LangtoFP}
In this appendix we start from the Langevin equation defined in Eqns. \ref{Langevin} and \ref{defLang} and derive the Fokker--Planck equation in Eqn.\ \ref{genbdfin}. As described in Sec.\ \ref{sec:Langevin} the observed trajectory is one realization $r_{\eta}(t)$ of the random stochastic process with Gaussian noise as presented in Eqn.\ \ref{trajnoise}:
\beq
p_n(t)=\int {\cal{D}} \eta  \delta \left(n-r_{\eta}(t) \right) P\left[\eta\right] := \avg{\delta \left(n-r_{\eta}(t) \right)}.
\eeq
Consider the evolution of the probability distribution:
\beq
p_n(t+\Delta t)-p_n(t)= \avg{\delta \left(n-r_{\eta}(t+\Delta t) \right)-\delta \left(n-r_{\eta}(t) \right)}, 
\eeq
and expand the increment in the trajectory as $r(t+\Delta t)= r(t)+ \Delta r(t)$. We can now Taylor expand the difference in delta functions:
\beqn
p_n(t+\Delta t)-p_n(t)&=& \avg{(-\Delta r(t))\delta' \left(n-r_{\eta}(t) \right)+\frac{1}{2} (-\Delta r(t))^2\delta'' \left(n-r_{\eta}(t) \right)}, \\
&=&-\partial_n \avg{\Delta r(t) \delta \left(n-r_{\eta}(t) \right)}+\frac{1}{2} \partial^2_n \avg{(\Delta r(t))^2 \delta \left(n-r_{\eta}(t) \right)},
\eeqn 
where the primes denote derivatives in $n$. Using the Langevin equation to calculate the increments in the trajectories:
\beq
\Delta r(t)=v\left[r(t)\right]\Delta t+\eta(t)\Delta t
\eeq
we obtain:
\beqn
\avg{\Delta r(t) \delta \left(n-r_{\eta}(t) \right)}&=& \avg{v\left[r(t)\right] \Delta t \delta \left(n-r_{\eta}(t) \right) }+ \avg{\eta(t) \Delta t \delta \left(n-r_{\eta}(t) \right) }\\
&&= v\left[r(t)\right] \Delta t \avg{\delta \left(n-r_{\eta}(t) \right) }= v\left(n \right])\Delta t  p_n(t)\\
 \avg{(\Delta r(t))^2 \delta \left(n-r_{\eta}(t) \right)}&=& (\Delta t)^2 \avg{(v\left[r(t)\right] )^2 \delta \left(n-r_{\eta}(t) \right) }+2 \avg{v\left[r(t)\right] \eta(t) \delta \left(n-r_{\eta}(t) \right) }+\avg{\eta^2(t) \delta \left(n-r_{\eta}(t) \right) }\\
 &&=  (\Delta t)^2 \left( v(n)\right)^2p_n(t)+ \avg{\eta^2(t) } p_n(t),
\eeqn
where we have used (assuming a discretized process):
\beq
\avg{\eta(t) \delta \left(n-r_{\eta}(t) \right)}= \avg{\eta(t)}\avg{\delta \left(n-r_{\eta}(t) \right)}=0.
\eeq
We identify 
\beq
D(n)=\frac{\Delta t}{2}\avg{\eta^2(t) }.
\eeq
Putting together all elements and keeping only leading terms in $\Delta t$:
\beq
\frac{p_n(t+\Delta t)-p_n(t)}{\Delta t}= -  \partial_n  \left[v\left(n \right])\Delta t  p_n(t) \right]+  \partial^2_n  \left[  D(n)  p_n(t) \right].
\eeq
Taking the limit of  $\Delta t \rightarrow 0$ we recover the Fokker--Planck equation:
\beq
\partial_t p_n(t)=  -  \partial_n  \left[v\left(n \right])\Delta t  p_n(t) \right]+  \partial^2_n  \left[  D(n)  p_n(t) \right].
\eeq

\section{Derivation of the Hill function}\label{Hillderiv}
The Hill function can be derived as an effective production rate for an autoregulating gene with two production states (Sec.\ \ref{burstauto}).  The gene is found either in an inactive state ($-$), in which the production rate is a constant $g_-$, or in an active state ($+$), in which the production rate $\omega_+n^h$ depends on the number of proteins $n$, which incorporates the autoregulation; $h$ describes the cooperativity, with $h>0$ corresponding to activation and $h<0$ corresponding to repression.
The probability $p_n^\pm$ of the gene being in a given state $+$ or $-$ and there being $n$ proteins evolves in time according to the master equation in Eqn. \ref{burstME} with $\Omega_{zz'}$ as in Eqn.\ \ref{Omega2n}; in steady state:
\beqn
\label{HillderME}
0&=&-{\cal L}^\pm p^\pm_n \pm \omega_+ n^{h} p^-_n \mp  \omega_-  p^+_n, 
\eeqn
where ${\cal L}^\pm$ describe simple birth--death terms with constant production rates $g_\pm$ in each of the two states.
We define moments of the master equation as
\beqn
\label{mean0Hllder}
\pi^\pm &\equiv& \sum_n p^\pm_n,\\ 
\label{mean1Hllder}
\pi^\pm \mu_\l^\pm &\equiv& \sum_n p^\pm_n n^\l \quad {\rm for} \,\,\,\l \ge 1.
\eeqn
with
\beq
\label{pidefHillder}
\pi^+ + \pi^- = 1
\eeq
by normalization.

Summing Eqn. \ref{HillderME} (top signs) over $n$ (and recalling that the birth--death terms sum to zero) gives
\beqn
0 &=& \omega_+ \sum_n n^h p^-_n -  \omega_-  \sum_n p^+_n \\
\label{sumrelation}
&=& \omega_+ \pi^- \mu_h^- -  \omega_- \pi^+
\eeqn
which, with Eqn.\ \ref{pidefHillder}, becomes an expression for $\pi^+$, the probability of being in the active state:
\beq
\label{partHillder}
 \pi^+= \frac{\mu_h^-}{\mu_h^- + \omega_+/\omega_-}.
\eeq
We solve for $\mu_h^-$ using two approximations.  The first is that higher moments can be be decoupled, i.e.\
\beq
\mu_{h+1}^- \approx \mu_h^- \mu_1^- \quad {\rm for} \,\,\,h \ge 1,
\eeq
which implies that
\beq
\label{decoupled}
\mu_h^- \approx (\mu_1^-)^h.
\eeq
This approximation allows one to simplify the mean equation in the $+$ state (obtained by summing the Eqn.\ \ref{HillderME}, top signs, against $n$ over $n$):
\beqn
0 &=& \pi^+ g_+ - \sum_n n p_n^+ + \omega_+ \sum_n n^{h+1} p_n^- - \omega_- \sum_n n p_n^+ \\
&=& \pi^+ g_+ - \pi^+\mu_1^+ + \omega_+ \pi^- \mu_{h+1}^- - \omega_- \pi^+ \mu_1^+ \\
&\approx& \pi^+ g_+ - \pi^+\mu_1^+ + \omega_+ \pi^- \mu_{h}^- \mu_1^- - \omega_- \pi^+ \mu_1^+ \\
\label{meanrelationfinal}
&=& \pi^+ g_+ - \pi^+\mu_1^+ + \omega_- \pi^+ \left( \mu_1^- - \mu_1^+ \right),
\eeqn
where the last step uses Eqn.\ \ref{sumrelation}.  The second approximation is that transitions between states are fast, i.e.\ $\omega_+ \sim \omega_- \gg 1$.  This approximation allows us to neglect the first two terms of Eqn.\ \ref{meanrelationfinal} compared to the third term, which implies that $\mu_1^- \approx \mu_1^+$.  Summing Eqn.\ \ref{mean1Hllder} over $\pm$ for $\l = 1$ then gives
\beq
\pi^+\mu_1^+ + \pi_- \mu_1^- \approx \left( \pi^+ + \pi^- \right) \mu_1^- = \mu_1^- = \sum_n p_n n = \bar{n},
\eeq
which shows that $\mu_1^-$ approximates the mean of the distribution.  Therefore with Eqn.\ \ref{decoupled} the probability that the gene is active (Eqn.\ \ref{partHillder}) can be written
\beq
\pi^+= \frac{\bar{n}^h}{\bar{n}^h + K},
\eeq
with equilibrium constant $K \equiv \omega_+/\omega_-$.  The effective production rate is the sum of the production rates in each state times the corresponding probabilities of being in each state:
\beqn
g(\bar{n}) &=& g_- \pi^- + g_+ \pi^+ \\
&=& g_- \left( 1 - \frac{\bar{n}^h}{\bar{n}^h + K} \right) + g_+ \frac{\bar{n}^h}{\bar{n}^h + K} \\
&=&  \frac{g_-K + g_+\bar{n}^h}{\bar{n}^h + K},
\eeqn
which is the Hill function (Eqn.\ \ref{regfun}).

\section{Limiting case of the two-state gene} \label{app:sri}

Here we show that the spectral solution to transcriptional bursting (Eqn.\ \ref{css}) reduces to the hypergeometric form (Eqn.\ \ref{Gpm}) in the limit of $Z=2$ production states.  We also derive the slightly simpler expression for the special case of zero production in the inactive state.

In the case of two states (Eqns.\ \ref{ppm}-\ref{Omega2}; $z = \pm$), Eqn.\ \ref{css} becomes
\beq
jG_j^\pm+\omega_\mp G_j^\pm-\omega_\pm G_j^\mp = \omega_\pm\sum_{j'<j}G_{j'}^\mp\frac{(\mp\Delta)^{j-j'}}{(j-j')!},
\eeq
where $\Delta = \Delta_{+-} = -\Delta_{-+}$.  Initializing with $G_0^\pm = \omega_\pm/(\omega_++\omega_-)$ and computing the first few terms reveals the pattern
\beqn
G_j^\pm &=& \frac{\omega_\pm}{\omega_++\omega_-}\frac{(\mp\Delta)^j}{j!}\frac{\prod_{j'=0}^{j-1}(j'+\omega_\mp)}{\prod_{j''=0}^{j-1}(j''+\omega_++\omega_-+1)} \\
&=& \frac{\omega_\pm}{\omega_++\omega_-}\frac{(\mp\Delta)^j}{j!}\frac{\Gamma(j+\omega_\mp)}{\Gamma(\omega_\mp)}\frac{\Gamma(\omega_++\omega_-+1)}{\Gamma(j+\omega_++\omega_-+1)},
\eeqn
where in the second line the products are written in terms of the Gamma function.
Writing the total generating function $\ket{G}=\sum_\pm\ket{G_\pm}$ in position space recovers the hypergeometric form (Eqn.\ \ref{Gpm}):
\beqn
G(x) &=& \sum_\pm\ip{x}{G_\pm} \\
&=& \sum_\pm\sum_j\ip{x}{j_\pm}\ip{j_\pm}{G_\pm}\\
&=& \sum_\pm\sum_j(x-1)^j e^{g_\pm(x-1)}G_j^\pm\\
\label{Ghyp}
&=& \sum_\pm \frac{\omega_\pm}{\omega_++\omega_-}e^{g_\pm(x-1)}
	\Phi[\omega_\mp,\omega_++\omega_-+1;\mp\Delta(x-1)],
\eeqn
where
\beq
\label{hyp2}
\Phi[\alpha,\beta;u] = \sum_{j=0}^\infty \frac{\Gamma(j+\alpha)}{\Gamma(\alpha)}\frac{\Gamma(\beta)}{\Gamma(j+\beta)}\frac{u^j}{j!}
\eeq
is the confluent hypergeometric function of the first kind.

In the limit $g_-=0$, Eqn.\ \ref{Ghyp} reads
\beq
\label{Ghyp2}
G(x)=\frac{\omega_+}{\omega_++\omega_-}e^u
	\Phi[\omega_-,\omega_++\omega_-+1;-u]
	+\frac{\omega_-}{\omega_++\omega_-}
	\Phi[\omega_+,\omega_++\omega_-+1;u],
\eeq
where $u\equiv g_+(x-1)$.  Using the fact that \cite{Koepf}
\beq
e^u\Phi[\alpha,\beta;-u] = \Phi[\beta-\alpha,\beta;u],
\eeq
Eqn.\ \ref{Ghyp2} can be written
\beq
G(x)=\frac{\omega_+}{\omega_++\omega_-}
	\Phi[\omega_++1,\omega_++\omega_-+1;u]
+\frac{\omega_-}{\omega_++\omega_-}
	\Phi[\omega_+,\omega_++\omega_-+1;u],
\eeq
or, noting Eqn. \ref{hyp2} and the fact that $\Gamma(s+1)=s\Gamma(s)$ for any $s$,
\beqn
G(x)&=&\sum_j\left(\frac{\omega_+}{\omega_++\omega_-}\frac{\Gamma(j+\omega_++1)}{\Gamma(\omega_++1)}
+\frac{\omega_-}{\omega_++\omega_-}\frac{\Gamma(j+\omega_+)}{\Gamma(\omega_+)}\right)
\frac{\Gamma(\omega_++\omega_-+1)}{\Gamma(j+\omega_++\omega_-+1)}\frac{u^j}{j!},\\
&=&\sum_j\left(\frac{\omega_+}{\omega_++\omega_-}\frac{(j+\omega_+)\Gamma(j+\omega_+)}{\omega_+\Gamma(\omega_+)}
+\frac{\omega_-}{\omega_++\omega_-}\frac{\Gamma(j+\omega_+)}{\Gamma(\omega_+)}\right)
\frac{(\omega_++\omega_-)\Gamma(\omega_++\omega_-)}{(j+\omega_++\omega_-)\Gamma(j+\omega_++\omega_-)}\frac{u^j}{j!},\qquad\\
&=&\sum_j\frac{\Gamma(j+\omega_+)}{\Gamma(\omega_+)}\frac{\Gamma(\omega_++\omega_-)}{\Gamma(j+\omega_++\omega_-)}\frac{u^j}{j!}\\
\label{sri2}
&=&\Phi[\omega_+,\omega_++\omega_-;u],
\eeqn
as in Eqn.\ \ref{sri}.
The marginal $p_n$ is obtained by $p_n = \partial_x^n[G(x)]_{0}/n!$;
using Eqn.\ \ref{sri2} and the derivative of the confluent hypergeometric function,
\beq
\partial_u^n\Phi[\alpha,\beta;u] = \frac{\Gamma(n+\alpha)}{\Gamma(\alpha)}\frac{\Gamma(\beta)}{\Gamma(n+\beta)}\Phi[\alpha+n,\beta+n;u],
\eeq
one obtains
\beqn
\label{raj2}
p_n &=& \frac{g_+^n}{n!}\frac{\Gamma(n+\omega_+)}{\Gamma(\omega_+)}\frac{\Gamma(\omega_++\omega_-)}{\Gamma(n+\omega_++\omega_-)}
\Phi[\omega_++n,\omega_++\omega_-+n;-g_+],
\eeqn
as in Eqn.\ \ref{raj}.

\bibliography{bode}

\end{document}